\documentclass[twocolumn,superscriptaddress,amsmath,amssymb,aps,prb,floatfix]{revtex4-2}

\usepackage{graphicx}
\usepackage{bm}
\usepackage{physics}
\usepackage[colorlinks,
    linkcolor=blue,
    anchorcolor=red,
    citecolor=green
]{hyperref}
\usepackage{wasysym}
\usepackage{MnSymbol}
\usepackage{xcolor}
\usepackage[thinc]{esdiff}
\usepackage{caption}
\usepackage[normalem]{ulem}

\newcommand{\PRLSection}[1]{\noindent {\color{blue} \textbf{ #1}}}

\usepackage{wasysym}

\begin{document}
\title{Quantum Restored Symmetry Protected Topological Phases}
\author{Dhruv Tiwari}
\affiliation{%
Max-Planck Institute for Solid State Research, 70569 Stuttgart, Germany
}%
\affiliation{%
Max Planck Institute for the Physics of Complex Systems, 01187 Dresden, Germany
}
\author{Steffen Bollmann}
\affiliation{%
Max-Planck Institute for Solid State Research, 70569 Stuttgart, Germany
}%

\author{Sebastian Paeckel}
\affiliation{%
Department of Physics, Arnold Sommerfeld Center for Theoretical Physics (ASC), Munich Center for Quantum Science and Technology (MCQST), Ludwig-Maximilians-Universit\"at M\"unchen, 80333 M\"unchen, Germany
}%

\author{Elio J. K\"onig}
\affiliation{%
Max-Planck Institute for Solid State Research, 70569 Stuttgart, Germany
}%
\affiliation{%
Department of Physics, University of Wisconsin-Madison, Madison, Wisconsin 53706, USA}

\begin{abstract}

Symmetry protected topological (SPT) phases are fundamental quantum many-body states of matter beyond Landau's paradigm. 
Here, we introduce the concept of quantum restored SPTs (QRSPTs), where the protecting symmetry is spontaneously broken at each instance in time, but restored after time average over quantum fluctuations, so that topological features re-emerge. 
To illustrate the concept, we study a one-dimensional fermionic Su-Schrieffer-Heeger model with fluctuating superconducting order. 
We solve this problem in several limiting cases using a variety of analytical methods and compare them to numerical (density matrix renormalization group) simulations which are valid throughout the parameter regime. We thereby map out the phase diagram and identify a QRSPT phase with topological features which are reminiscent from (but not identical to) the topology of the underlying free fermion system. 
The QRSPT paradigm thereby stimulates a new perspective for the constructive design of novel topological quantum many-body phases. 

\end{abstract}

\maketitle

Symmetries play an exceptional role in characterizing quantum materials. On the one hand, following Landau's legacy~\cite{Landau1937}, spontaneous symmetry breaking (SSB) has been of paramount importance for classifying many-body ground states. More recent advances demonstrate that, even when symmetry breaking order parameters form locally, strong quantum fluctuations of their orientation may impede true SSB and give way to exotic phenomena such as vestigial order~\cite{fernandesSchmalian2019} and quantum paramagnetism such as quantum spin liquids~\cite{SavaryBalents2016}. On the other hand, symmetries are also of paramount importance for characterizing order beyond the Landau paradigm. Specifically, a given Symmetry Protected Topological (SPT) phase~\cite{ChenWen2013,Senthil2015,Kitaev2013video} represents a class of gapped short-range entangled many-body quantum states that cannot be connected adiabatically to a different SPT phase as long as the symmetry is unbroken~\cite{ChenWen2010}.
Amongst the most glaring properties that are robust to symmetry-preserving perturbations are protected gapless boundary excitations in the presence of non-trivial SPT order~\footnote{There exist examples of Quotient Symmetry Protected Topological phases where the underlying protecting symmetry is a quotient of the original symmetry group (of the underlying Hamiltonian) and thus it is possible to adiabatically connect the trivial and non-trivial SPT phases even if the underlying symmetry is unbroken ~\cite{VerresenPollmann2021}. Such phases tend to exhibit a boundary transition without a corresponding bulk transition.}. 

\begin{figure}
    \includegraphics[scale = 1]{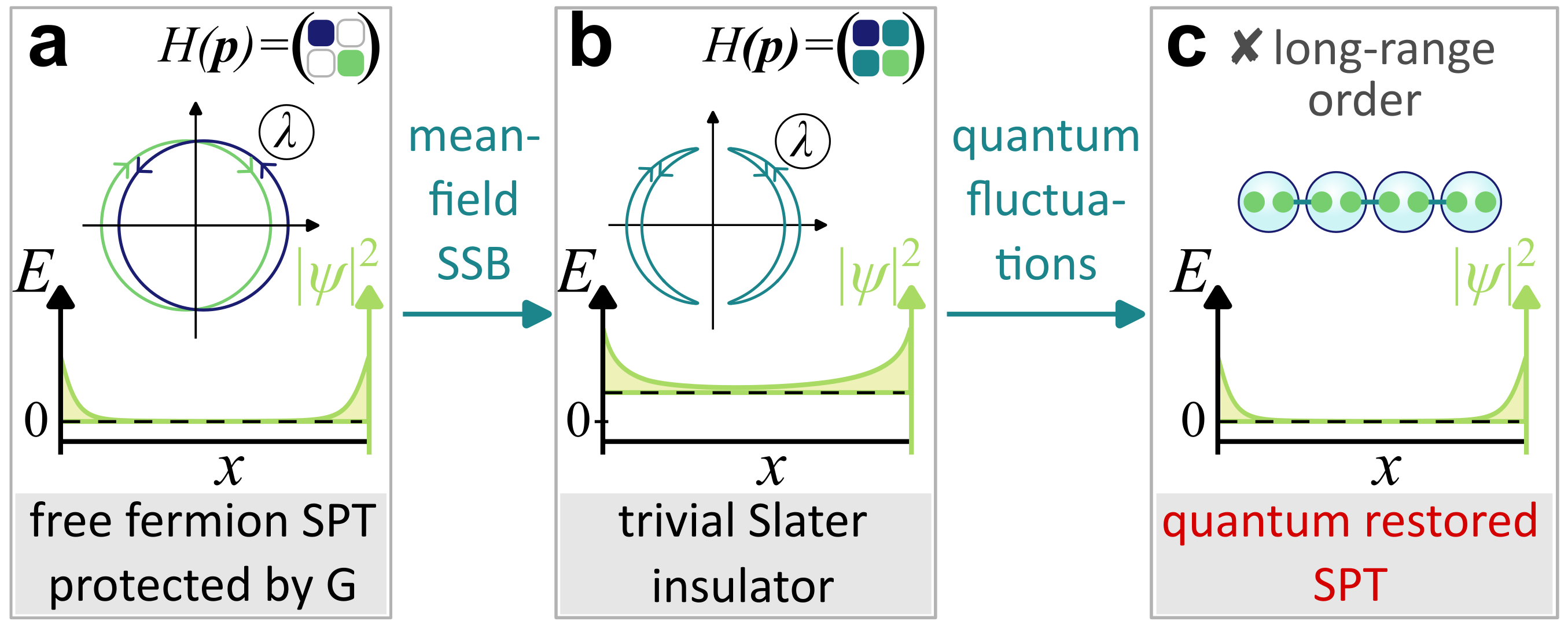}
    \caption{a) Consider a free fermion SPT where the symmetry group G prevents the admixture of distinct topological sectors of the Bloch Hamiltonian and of zero energy boundary states with distinct quantum numbers. b) A mean-field SSB of G trivializes such a quantum state, but c) strong quantum fluctuations of the order parameter may destroy long-range order even if the local expectation value of the order parameter amplitude is non-zero. Thereby, topological features reemerge, and the SPT is quantum restored.}
    \label{Fig:QRSPT}
\end{figure}

Classic examples of states exhibiting non-trivial SPT order are free fermion topological insulators~\cite{QiZhang2011} and the bosonic, strongly correlated ``Haldane'' phase of antiferromagnetic spin-1 chains~\cite{Haldane1983,AffleckTasaki1987}.
There exist multiple theoretical and numerical ways of characterizing and detecting phases (or classes) of SPT order~\cite{ChenWen2010, FidkowskiKitaev2011, PollmannOshikawa2010, TurnerBerg2011, PollmannTurner2012, GuWen2014, RyuLudwig2010, Kitaev2009, HaegmanSchuch2012} in particular, 
the classification of interacting SPT phases (for a given symmetry group) can be different from their non-interacting counterpart~\cite{FidkowskiKitaev2010, LapaHughes2016}. 

Of particular interest for the present work are SPTs where the protecting symmetry is present only on average. The historically first example regards disordered systems where the symmetry may be broken in each realization, but restored upon ensemble average~\cite{OstrovskyMirlin2007,FuKane2012,KoenigMirlin2014,MaWang2023}. In mathematical terms, consider disorder fields $\phi$ with a probability distribution $\mathcal P[\phi]$, which is invariant under the protecting symmetry. Then, the effective Euclidean action of replicated matter fields $\psi_r$ 
\begin{equation}
    S_{\rm eff}[\{\psi_r\}] = - \text{ln}\left (\int \mathcal D\phi \mathcal P[\phi] \exp\{- \sum_r S_0[\psi_r,\phi]\} \right ),
\end{equation}
may display SPT features even if $S_0[\psi,\phi]$ for a given configuration $\phi$ breaks the protecting symmetry. Recent advances generalize the concept of such average SPTs to amorphous systems~\cite{Grushin2023} and open quantum systems with decoherence and mixed quantum states~\cite{MaChong2023}.

In this paper we introduce the concept of quantum restored symmetry protected topological (QRSPT) phases:  They occur in failed SSB states where strong quantum fluctuations of the local order parameter orientation lead to restoration of a symmetry protecting an SPT. Subsequently, the main part of the paper is devoted to an exemplary model, a spinful Su-Schrieffer-Heeger model with fluctuating s-wave superconductivity, displaying the outlined general phenomenology. We first provide a summary of the results, which we subsequently derive carefully using a variety of analytical and numerical methods.

\PRLSection{What are quantum restored SPTs?} 
For an illustration of the general QRSPT paradigm, see Fig.~\ref{Fig:QRSPT}. First, concentrate on a clean, free fermion topological insulator or superconductor with symmetry group $G$. Unitary symmetries allow to block diagonalize the Bloch Hamiltonian, each block being subsequently topologically scrutinized by the tenfold-way methodology~\cite{SchnyderLudwig2008}. In particular, we consider the case where the topological invariants in the various blocks are non-zero but sum up to zero. Importantly, the protecting symmetry ensures different quantum numbers for zero boundary states emanating from different blocks precluding any mutual annihilation.

Next, Fig.~\ref{Fig:QRSPT} b), we include interactions and assume that the symmetry which ensures the block-diagonalization of the Hamiltonian is spontaneously broken at the mean-field level. One may still resort to non-interacting band-topology but the Bloch Hamiltonian can no longer be block diagonalized. It must thus be treated as a whole and the topological invariant vanishes. 
The formation of mean-field order parameters is insufficient to demonstrate actual SSB. In particular, for continuous groups $G$, quantum fluctuations of the orientation of the order parameter (``Goldstone modes'' $\phi$) may inhibit true long-range order. In this case, at the longest time/length scales $G$ symmetry is recovered and can lead to quantum restoration of SPT phenomenology, Fig.~\ref{Fig:QRSPT} c). In mathematical terms, the effective Euclidean action of fermions $\psi$
\begin{equation}
    S_{\rm eff}[\psi] = - \text{ln}\left (\int \mathcal D\phi \, \exp\{-  S[\psi,\phi] - S_{\rm Goldstone}[\phi]\} \right ), \label{eq:QRSPT}
\end{equation}
displays SPT features, even though each given order parameter field configuration $\phi$ breaks the protecting symmetry $G$.

We conclude this section with a few comments.
First, in the above we discussed the situation of intrinsic SSB within the topological fermionic system. In one and two dimensions it is equally conceivable that the fluctuating order parameter and the Goldstone modes $\phi$ emanate from proximitizing the $G$-symmetric fermionic material with a second material with spontaneous symmetry breaking (e.g.~a magnet or superconductor). Such a heterostructure also represents bona-fide implementation of two-fluid models of the type of Eq.~\eqref{eq:QRSPT}.
Second, we highlight that the interplay of Dirac electrons with quantum disordered order parameters, in particular through condensation of topological defects, has been discussed in the past particularly in connection with interacting topological insulator boundary states~\cite{WangSenthil2014} and with the paradigm of symmetric mass generation~\cite{WangYou2022}. Third, one may wonder what happens to the system if the quantum fluctuations are weak and the true-long range order is established. Per Goldstone's theorem, the bulk system is gapless, yet it is possible that the underlying free fermion topology enforces the emergence of additional topological terms in the action describing order parameter fluctuations~\cite{RamppSchmalian2022,BollmannKoenig2023b} leading potentially to topological Goldstone phases of matter~\cite{Else2021}.
Finally, one may argue that $S_{\rm eff} [\psi]$ is nothing but a model for a very specific interacting fermionic SPT~\cite{Gurarie2011,GuWen2014,YouXu2014}, in particular when $\phi$ correlations are short-range deep in the quantum disordered state, so that $S_{\rm eff} [\psi]$ describes a local theory. While this statement is in principle true, the model scrutinized in this paper demonstrates that the QRSPT paradigm promises much richer physics, in particular near the phase transitions of the system, where fermions and Goldstone bosons mutually stabilize the physics of long-range interacting quantum systems~\cite{DefenuTrombettoni2023} (the classic solid state example are RKKY~\cite{RudermannKittel1954,Kasuya1956,Yosida1957} interactions).

\begin{figure}
     \centering
    \includegraphics[scale = 1]{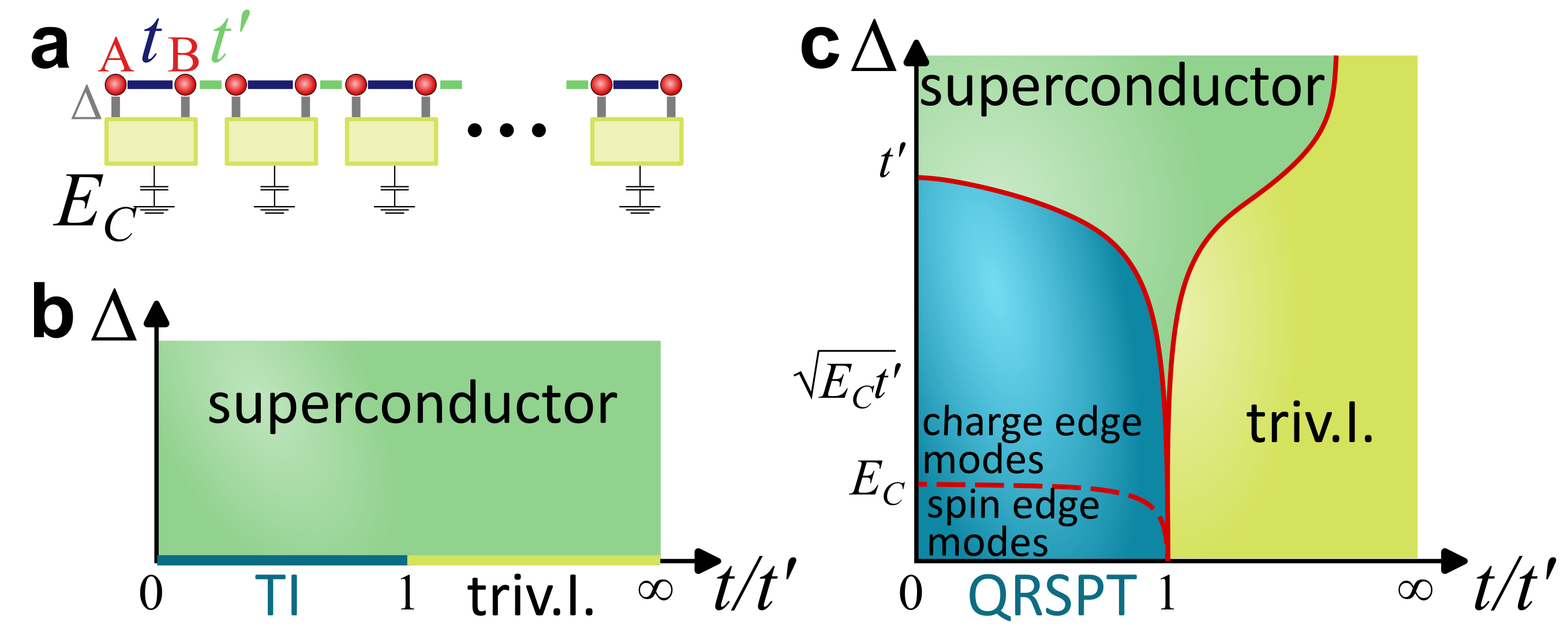}
    \caption{a) Schematic representation of the model Eq.~\eqref{Eq: Hamiltonian for the total model.}. The rectangular boxes denote the Cooper pair boxes. b) Mean-field phase diagram: $\Delta$ breaks the symmetry protecting the free-fermion topology (cf Fig.~\ref{Fig:QRSPT} b)) so that topological and trivial phase may be adiabatically connected. TI and {triv.I.} represent topological and trivial insulator respectively. Note that there is a gap closure (as expected from standard SSH physics) at $t=t^{\prime}$ (for $\Delta = 0$). c) Schematic phase diagram of Eq.~\eqref{Eq: Hamiltonian for the total model.}. Note the QRSPT phase at small $\Delta$ and  $t < t'$.}
    \label{Fig:Model}
\end{figure}

\PRLSection{Model:} As a paradigmatic model to illustrate the QRSPT concept we study a mesoscopic topological Josephson junction array.
We assume fermionic quantum dots forming a spinful Su-Schrieffer-Heeger (SSH) \cite{SuHeeger1980,RyuHatsugai2006} chain, Fig.~\ref{Fig:Model} a) and couple it to an array of floating superconducting islands \cite{LiVayrynen2023} as follows

\begin{align}
   \hat{H} &= E_{\rm{C}}\sum_{X} (2\hat{N}_{X} + \hat{n}_{X} - N_{g})^{2} \notag \\
   &-\sum_{X,\sigma} (t \, d^{\dagger}_{X,A,\sigma}d_{X,B,\sigma} + t^{\prime} \, d^{\dagger}_{X+1,A,\sigma}d_{X,B,\sigma} + \mathrm{H.c.}) \notag \\
   &-\frac{\Delta}{2}\sum_{X,j,\sigma,\sigma^{\prime}} (e^{-i\hat{\phi}_{X}}d^{\dagger}_{X,j,\sigma}[\sigma_{y}]_{\sigma,\sigma^{\prime}}d^{\dagger}_{X,j,\sigma^{\prime}} + \mathrm{H.c.}),
\label{Eq: Hamiltonian for the total model.}
\end{align}
where $X \in \mathbb Z$, $j \in$ $\{A,B\}$ and $\sigma \in \{\uparrow, \downarrow \}$ denote the unit cell, sublattice of SSH chain and the spin, respectively. The operator $\hat{N}_{X} = -i \partial_{\hat{\phi}_X}$ measures the number of Cooper-pairs at the $X$th Cooper-pair box and is conjugate to the Cooper-pair annihilator $e^{-i\hat{\phi}_{X}}$(i.e. to the fluctuating superconducting order parameter). Similarly $\hat{n}_{X} = \sum_{j,\sigma} d^\dagger_{X,j,\sigma}d_{X,j,\sigma}$ measures the number of electrons in the unit cell $X$. We denote the unit cell using the uppercase index $X$ whereas lowercase index $x$ represents the continuum position variable (appears later in the paper). 
Note that the proximity-induced coupling strength $\Delta$ is much smaller than the bulk superconducting gap, this allows to ignore the quasiparticle states within the superconductor for physical considerations limited to the lowest energy excitations of the model. We also assume that the dimensions of the superconductor are much larger than the coherence length of the Cooper-pairs, this allows us to neglect crossed Andreev reflection
~\cite{LiVayrynen2023, KnappLutchyn2020,ShiKarsten2020}. For simplicity, we set the Josephson coupling between the superconducting islands to be zero. 
We restrict ourselves to $E_{\rm{C}} < \sqrt{t^{2} + t^{\prime 2}}$ (since our analytical calculations are only valid in this regime) and even values of the gate voltage $N_{g}$. 
The basic symmetries of Eq.~\eqref{Eq: Hamiltonian for the total model.} are the conservation of total charge $\sum_X 2\hat N_X + \hat n_X$ ($U(1)$ symmetry) and a combined sublattice and particle-hole transformation denoted by $C$, see methods and \cite{SuppMat}.

\PRLSection{Summary of results:} We first elucidate in which sense Eq.~\eqref{Eq: Hamiltonian for the total model.} displays QRSPT phenomenology. At $\Delta = E_{\rm{C}} = 0$, the fermionic sector decouples from bosons and displays standard SSH topology. Next, we include superconductivity within a mean-field approximation which corresponds to $E_{\rm{C}} = 0$. The superconducting phase at each island becomes a classical variable which we gauge to $\phi_X = 0,$ throughout. The mean-field Bloch Hamiltonian in Nambu space (see methods) ceases to be block diagonal in the presence of $\Delta$ ($U(1)$ and $C$ symmetry breaking) term, thereby losing the compensated topology of electron and hole bands, cf. the loss of winding numbers illustrated in Fig.~\ref{Fig:QRSPT} a) vs. b).
The model is gapped for all non-zero values of $\Delta$, Fig. \ref{Fig:Model} b), thus allowing for an adiabatic connection of the topological phase of the SSH model (at $\Delta 
 = 0, t<t^{\prime}$) to the trivial phase (at $\Delta 
 = 0, t>t^{\prime}$).
 Simultaneously, the edge spectrum of the SSH model is gapped out for non-zero values of $\Delta$ as the spinful fermionic edge states combine into Cooper pairs.
 Quantum fluctuations beyond the mean-field approximation are introduced by Coulomb interactions (represented by $E_{\rm{C}}$ term). 
The phase diagram, schematically shown in Fig.~\ref{Fig:Model} c), contains a QRSPT phase at small $\Delta$ and $t<t'$, a superconductor emanating from the free fermion critical point $\Delta = 0, t = t'$ and a trivial gapped phase for $t>t'$ and small $\Delta$.
Leaving details to the remainder of the paper, we now summarize the pecularities of these phases. Most importantly, the QRSPT phase is characterized by edge states even for non-zero values of $\Delta$ as derived by perturbing the system around an integrable limit at $t \ll t'$ and by analyzing soliton solutions of the field theory near $t = t'$. A particular curiosity of the present model is the boundary transition from spin to charge edge modes within the QRSPT which is also confirmed numerically. Additionally, the topological nature is corroborated by symmetry fractionalization arguments and, numerically, by the observation of degeneracies in the entanglement spectrum. Unlike the mean-field analysis, we do observe both in analytical field theory and DMRG a phase transition (or intermediate phase) near $t = t'$ for non-zero values of $\Delta$. It separates the topological phase from a trivial phase without edge states or degeneracies in the entanglement spectrum. The critical point serves as a seed for a superconducting phase at large $\Delta$, which is stabilized by emergent Josephson coupling between the islands. 

\PRLSection{Perturbation about the dimerized limits:} We first study the effect of introducing  $\Delta$ and $E_{\rm{C}}$ perturbatively about the two extreme regimes corresponding to $t=0$ and $t^{\prime} = 0$ assuming $\Delta,E_{\rm{C}} \ll \sqrt{t^{2} + t^{\prime 2}}$. In the former regime, contrary to the mean-field results, we do observe a gapless edge spectrum (accompanied by a peculiar boundary transition from gapless spin edge modes to gapless charge edge modes) for $\Delta \neq 0$ as discussed below.\\
For $t = 0$ ($\Delta, E_{\rm{C}} = 0$), the SSH model is in its topological state with intercell dimers. Assuming periodic boundary conditions, the groundstate of the SSH model in this regime is given by:
\begin{small}
\begin{align}
        &\ket{\psi (\lbrace N_{\rm{X}}\rbrace)} = \prod_{X=1}^{N} \ket{N_{X}} \otimes \ket{\psi_{\rm{SSH}}}, \label{Eq: Groundstate to perturb about for t=0}\\
        &\ket{\psi_{\rm{SSH}}} = \bigg[\prod_{X=1}^{N} \bigg(\frac{d^{\dagger}_{A,X,\uparrow} + d^{\dagger}_{B,X+1,\uparrow}}{\sqrt{2}}\bigg)\bigg(\frac{d^{\dagger}_{A,X,\downarrow} + d^{\dagger}_{B,X+1,\downarrow}}{\sqrt{2}}\bigg)\bigg] \ket{0} \notag,
\end{align}
\end{small}
where $N$ is the number of unit cells and $\ket{0}$ is the vacuum state. Note that the number $N_X$ of bosons in each cell is arbitrary in Eq.~\eqref{Eq: Groundstate to perturb about for t=0} since the terms $E_{\rm{C}}$ and $\Delta$ are zero implying no coupling of bosons and fermions. This massive degeneracy is lifted upon introducing $E_{\rm{C}}$ as a perturbation ($\Delta$ induced matrix elements within the ground state manifold vanish) so that the correction to groundstate energy is positive and extensive in $E_{\rm{C}}$ up to first orders in $\Delta$ and $E_{\rm{C}}$. The average fermion occupation of two in $\ket{\psi_{\rm SSH}}$ implies that the groundstate corresponds to Eq.~\eqref{Eq: Groundstate to perturb about for t=0} with $N_{X} = N_g-2$ ($N_g$ $\in$ $2\mathbb{Z}$),
see \cite{SuppMat} for details (we choose $N_g =2$ henceforth).

\begin{figure}
    \centering
    \includegraphics[scale = 1]{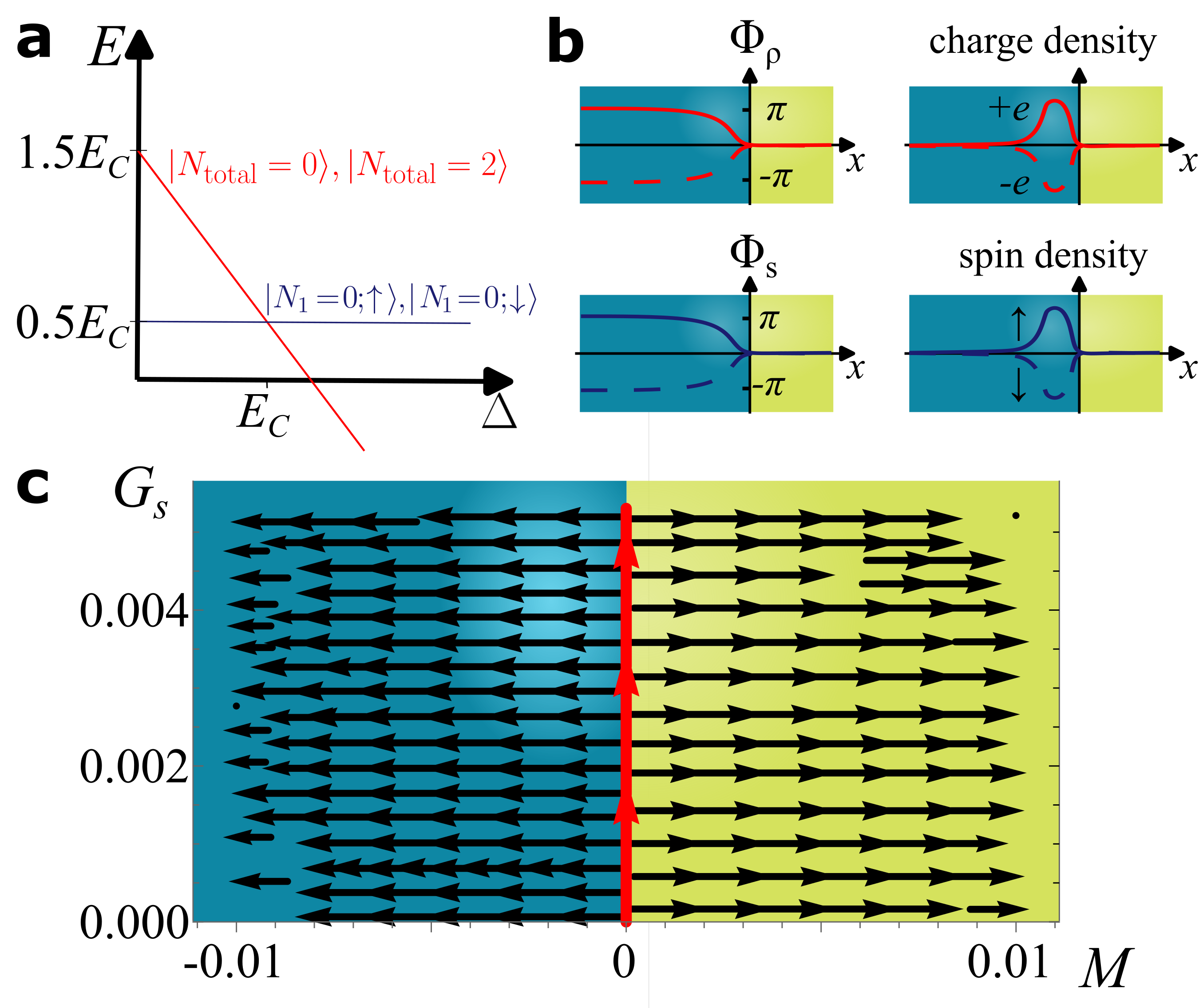}
    \caption{a) Eigenvalues of the edge Hamiltonian, Eq.~\eqref{Eq: Effective edge Hamiltonian}, as a function of $\Delta$ for $t=0, N_{g}=2$. Note the edge transition from gapless spin edge modes to gapless charge edge modes at $\Delta = E_{\rm{C}}$. $N_{1}$ represents the number of bosons on the first bosonic site. $N_{\rm{total}} = 2N_{1} + n_{1}$ represents the total charge on the left edge. b) Edge states correspond to kink-like field configurations within the effective field theory near the free fermion critical point, Eq.~\eqref{Eq: Bosonized Equation}.
    c) RG flow obtained using the flow equations given in Eq:\ref{Eq: RG flow equations}.}
    \label{Fig: Analytical results}
\end{figure}

To study the edge states of this model in this regime, we  transform to open boundary conditions (OBC) by assuming that the state $\ket{\psi_0} = \ket{\psi(\{N_{X}=0 \})}$ still describes the groundstate under OBC except for the first and last Cooper-pair box and fermionic site. 
Thus, the effective edge Hamiltonian is given by:
\begin{gather}
    \begin{aligned}
        \bra{\psi_0}\hat{H}\ket{\psi_0} &= \hat{H}_{\rm{left-edge}} + \hat{H}_{\rm{right-edge}} + E_{\rm{bulk}},
    \end{aligned}
\end{gather}
where $E_{\rm{bulk}}$ corresponds to the energy of the state $\ket{\psi_0}$ in the bulk.
The effective edge Hamiltonian on the left edge is given by:
\begin{gather}
    \begin{aligned}
    \hat{H}_{\rm{left-edge}} &= E_{\rm{C}} (2\hat{N}_{1} + \hat{n}_{A,1} -1)^2 \\
    &-\frac{\Delta}{2}(e^{-i\hat{\phi}_{1}}d^{\dagger}_{1,A}\sigma_{y}d^{\dagger}_{1,A} + \mathrm{H.c.}) + \frac{E_{\rm{C}}}{2}.
    \end{aligned}
\label{Eq: Effective edge Hamiltonian}
\end{gather}
This edge-Hamiltonian can be readily solved, the lowest energy edge excitations as a function of $\Delta$ are plotted in Fig.~\ref{Fig: Analytical results}(a). Analogous results hold at the right edge. While the ground state is always degenerate, an edge transition from gapless spin edge modes to gapless charge edge modes at $\Delta = E_{\rm{C}}$ occurs. 

We use techniques of symmetry fractionalization \cite{TurnerBerg2011,FidkowskiKitaev2011,TurnerVishwanath2013} to demonstrate that both the spin and charge edge modes are protected at least by the antiunitary particle-hole symmetry $\mathit{C}$, see \cite{SuppMat} for details. This antiunitary symmetry, which squares to unity in the bulk, fractionalizes at low energies into operators which act locally on the edge states. As the fractionalized representation squares to $-1$, it implies that $\mathit{C}$ protects edge degeneracy by a generalized Kramers theorem. Thus, contrary to the mean field results, there exists an SPT phase for non-zero values of $\Delta$. 
When a similar perturbative calculation is performed near $t^{\prime} = 0$, the edge spectrum is gapped throughout indicating a trivial phase and no correction to the groundstate energy occurs to first orders in $\Delta$ and $E_{C}$. The latter result highlights the fact that our model is asymmetrical upon an exchange of $t$ and $t^{\prime}$, see the schematic phase diagram shown in Fig.~\ref{Fig:Model} c).

\PRLSection{Field theory near the free fermion critical point:} We now turn to study signatures of the bulk phase transition for non-zero values of $\Delta$ and concentrate on the limit $\Delta \ll E_{\rm{C}}$, $\abs{t-t^{\prime}} \ll E_C \ll \sqrt{t^{2} + t^{\prime 2}}$.

For the unperturbed Hamiltonian (corresponds to $\Delta =0$ in Eq.~\eqref{Eq: Hamiltonian for the total model.}), the fermionic gap is controlled by the term $\abs{t-t^{\prime}}$ (the mass gap in the SSH model) and the bosonic gap is controlled by the term $E_{\rm{C}}$. Thus, bosons are fast as compared to fermions and can be integrated out to obtain an effective low-energy theory of interacting fermions, see~\cite{SuppMat}. The value of $\sqrt{t^{2} + t^{\prime 2}}$ serves as an estimate for the bandwidth of the single particle spectrum of free fermions (corresponding to $\Delta = E_{\rm{C}} = 0$) and the assumption of $E_{C} \ll \sqrt{t^{2} + t^{\prime 2}}$ controls the bosonization approach on top of the linearized fermionic Hamiltonian. Its bosonized representation is given by the following action:
\begin{gather}
    \begin{aligned}
        S &= \sum_{\alpha = s,\rho} \int_{x,\tau} \frac{1}{4 \pi K_{\alpha}} \bigg( \frac{(\partial_{\tau}\Phi_{\alpha})^{2}}{u_{\alpha}} + (\partial_{x}\Phi_{\alpha})^{2} u_{\alpha}  \bigg) \\
         &-\int_{x,\tau} [G_{s}\cos(2\Phi_{s}) + M \cos(\Phi_{s})\cos(\Phi_{\rho}) ],
    \end{aligned}
\label{Eq: Bosonized Equation}
\end{gather}
where $s$ and $\rho$ denote the spin and charge degrees of freedom respectively and $\Phi_{\rho,s} {\equiv} \Phi_{\rho,s} (x,\tau)$ are bosonic fields in the corresponding sector. In terms of parameters of Eq.~\eqref{Eq: Hamiltonian for the total model.}, $G_s \sim \Delta^2/(E_C a), M \sim (t'-t)/a$ while Luttinger parameters $K_{
\rho,s}$ and hydrodynamic velocities $u_{\rho,s}$ are contained in the methods section. To study the phase diagram of the effective action in Eq.~\eqref{Eq: Bosonized Equation} we perform a perturbative Renormalization Group (RG) \cite{Callan1970}  analysis of the bosonized action in Eq.~\eqref{Eq: Bosonized Equation} (about the fixed point of Luttinger liquid theory for both the charge and spin sector and up to first order in prefactors of cosines). The flow equations are
\begin{gather}
    \begin{aligned}
        \frac{dG_{s}}{dl} &= (2 - 2K_{s})G_{s}, \quad
        \frac{dM}{dl} &= (2 - \frac{K_{s}}{2} - \frac{K_{\rho}}{2})M,
    \end{aligned}
\label{Eq: RG flow equations}
\end{gather}
where $l = \ln(\Lambda/\Lambda')$ and $\Lambda$ ($\Lambda'$) is the momentum cutoff of the theory (the running momentum cutoff). Note that in the present theory $K_s <1$ hence the $G_s$ term is always relevant and the system displays a spin gap for any non-zero $\Delta$. In contrast, $K_\rho >1$ is possible and the mass term changes from relevant (displayed in Fig.~\ref{Fig: Analytical results} c)) to irrelevant (not shown) when $K_\rho + K_s = 4$ 
which corresponds to $\Delta \sim \sqrt{E_C t'}$ in terms of microscopic parameters entering Eq.~\eqref{Eq: Hamiltonian for the total model.}.

This leads to the phase diagram illustrated in Fig.~\ref{Fig:Model} c): For $\Delta \ll \sqrt{E_C t'}$, non-zero $M$ flow to a fully gapped and topologically trivial (non-trival) insulator, the two phases being divided by a critical line. In contrast, for $\Delta \gg \sqrt{E_C t'}$ one expects that the insulating phases are separated by a third phase without charge gap. Note that this requires pushing the field theory discussed in this section beyond its limits of applicability, hence we prove this claim later by complementary means. Importantly, unlike the mean-field result, a critical theory without a charge gap separates the two insulating phases even for non-zero values of $\Delta$. This gapless state corresponds to a singlet s-wave superconductor as it has dominant correlations of the type $e^{-i\frac{\Theta_{\rho}}{\sqrt{2\pi}}}\cos(\Phi_{s})$~\cite{Giamarchi2004,KeselmanBerg2015} where $\Theta_{\rho}$ is the bosonic field conjugate to $\Phi_{\rho}$.

We also studied edge modes at the interface of the phases corresponding to negative and positive values of $M$ using semiclassics~\cite{SuppMat}. Taking into account the periodicity of the bosonic variables $\Phi_{\rho}$ and $\Phi_{s}$, the groundstate in terms of the bosonic variables for $M > 0$ corresponds to $(\Phi_{\rho,i},\Phi_{s,i}) = (0,0)$. For $M < 0$, the groundstate corresponds to four possible values of the bosonic variables given by:
\begin{gather}
    \begin{aligned}
        (\Phi_{\rho,f},\Phi_{s,f}) &= (\pm \pi,0),(0,\pm\pi),
    \end{aligned}
\end{gather}
where these different configurations of ($\Phi_{\rho,f},\Phi_{s,f}$) lead to the same groundstate energy in the bulk but they imply four different edge states represented by half-kink or half-antikink in spin or charge sector, see Fig.~\ref{Fig: Analytical results} b). 
A spin mode leads to an accumulation of spin 1/2 at the edge which can be seen by calculating the corresponding 
 $S_{z}$ magnetization:
\begin{gather}
    \begin{aligned}
    S_{z} &= \frac{1}{2\pi} \int dx \partial_{x} \Phi_{s} 
    = \pm \frac12.
    \end{aligned}
\end{gather}
Similarly, charge modes lead to an edge accumulation of 
charge 
\begin{gather}
    \begin{aligned}
      N_{e}  &= \frac{e}{\pi} \int dx \partial_{x} \Phi_{\rho} 
      = \pm e,
    \end{aligned}
\end{gather}

relative of to the ground state charge configuration. We analytically determine~\cite{SuppMat} the kink energies demonstrating that the charge and spin kinks are each two-fold degenerate. Comparing those energies at non-zero $\Delta$, we find that the ground states within this effective continuum model feature spin (charge) edge states for small (large) $\Delta < \Delta_c$ ($\Delta > \Delta_c$) with a non analytic edge transition curve: 
\begin{equation}
    \Delta_{c} \simeq\sqrt{-\frac{0.177715 W(-\alpha \log (| \delta \theta | ))}{\log (| \delta \theta | )}},
\end{equation}
where $\delta \theta \simeq \frac{t-t^{\prime}}{\sqrt{t^{2} + t^{\prime2}}}$, $W$ represents Lambert's W function and $\alpha = 0.000034$. This curve is
is schematically shown in Fig.~\ref{Fig:Model} c), see \cite{SuppMat} for more details.
Far away from the bulk transition $\frac{t}{t^{\prime}} \ll 1$, the same formalism yields an edge transition at $\Delta \simeq E_{\rm{C}}$, consistent with the edge transition obtained from perturbation theory above.

\begin{figure*}
    \centering
    \includegraphics[scale = 1]{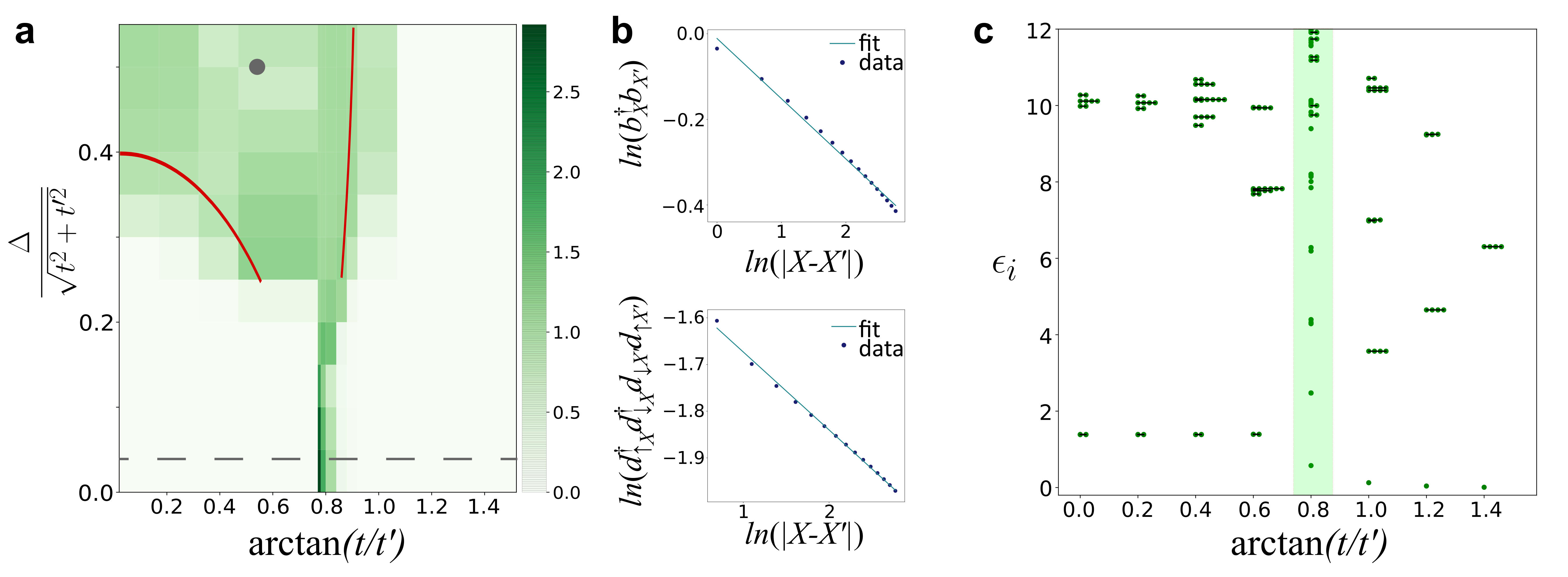}
    \caption{a) Color plot of the central charge obtained using finite DMRG for a system size of 40 and $N_{g} = 4$.
    The phase diagram is obtained for $\frac{E_{\rm{C}}}{\sqrt{t^{2}+ t^{\prime2}}} = 0.01$. Red lines denote the analytically obtained position of the Berezinskii-Kosterlitz-Thouless transition line. The grey dot (grey dashed line) correspond to the locations in parameter space at which the data in panels b) and c) are taken. b) Log-Log plot of the bosonic and fermionic correlator for $\Delta = 50 E_{\rm{C}}$ and $\arctan(\frac{t}{t^{\prime}}) = 0.47$. The power-law fit for the bosonic and fermionic correlator gives a $K_{\rm{SC}}$ value of $0.071 \pm 0.001$ and $0.070 \pm 0.001$ respectively.
    c) Entanglement spectrum obtained using iDMRG for $\Delta = 4 E_{\rm{C}}$ and $\frac{E_{\rm{C}}}{\sqrt{t^{2}+t^{\prime2}}} = 0.01$ and $N_{g} = 2$. The green-shaded regime corresponds to criticality. Consistently with analytical expectations, we observe even degeneracy throughout the regime corresponding to QRSPT in Fig.~\ref{Fig:Model}(c). Note that we only show the entanglement spectrum values upto $\epsilon_{i} = 12$.}
    \label{Fig:DMRG}
\end{figure*}

\PRLSection{Field theory perturbing about the superconducting phase.} In the regime  $E_{\rm{C}} < \Delta,\abs{t-t^{\prime}}$, the upper bound on the bosonic gap is controlled by the term $E_{\rm{C}}$ and is much smaller than the fermionic gap controlled by the term $\Delta$ and $\abs{t-t^{\prime}}$. Thus, we can integrate out the fermions in this regime and obtain a low energy effective theory of bosons. Around the regime where the bosons become gapless, the effective bosonic theory can be described by a Luttinger liquid and we determine corresponding Luttinger liquid parameters. The procedure of integrating out fermions is non-trivial for the total Hamiltonian in Eq.~\eqref{Eq: Hamiltonian for the total model.} due to the presence of the electron density $\hat{n}_{X}$ in the $E_{\rm{C}}$ term. To make the procedure relatively simpler and transparent, it is useful to perform a basis change to a basis where fermionic fields effectively follow the slowly fluctuating superconducting phase. 
The bosonic action obtained after integrating out fermions is given by
\begin{gather}
    \begin{aligned}
        S_{\rm{eff}} (\phi) &= \frac{1}{2K_{\rm{SC}}}\int_{x,\tau} \frac{{(
        \partial_\tau \phi)}^{2}}{u_{\rm{SC}}} + u_{\rm{SC}} (\partial_{x}\phi)^{2}.
    \label{eq:SCAction}
    \end{aligned}
\end{gather}
The main steps in the process of integrating out fermions and the definition of $K_{\rm{SC}}$ and $u_{\rm{SC}}$ in terms of model parameters are given in the methods section.
$K_{\rm{SC}}$ denotes the inverse superconducting stiffness and a phase transition from the insulating regime to a superconducting regime occurs at $K_{\rm{SC}}=1$. Note that this transition from the insulator to the superconductor is the Berezinskii-Kosterlitz-Thouless transition \cite{KosterlitzJames2018} and corresponds to the proliferation of phase slips. This condition and the expression $K_{\rm SC}$ in terms of microscopic parameters determines the phase boundaries plotted in red in Fig.~\ref{Fig:DMRG} a) (no fitting involved). On the side, we remark that within the effective Luttinger liquid theory, the superconducting regime is not observed for $\frac{E_{\rm{C}}}{\sqrt{t^{2} + t^{\prime2}}} > 0.02$, which could signal Mott localization throughout.

\PRLSection{DMRG results:} To model the Hamiltonian in Eq.~\eqref{Eq: Hamiltonian for the total model.} numerically, we truncate the local Hilbert space dimension of Cooper pairs (
created/annihilated by $e^{\pm i\hat{\phi}_{X}}$ 
to a finite value of 8 (we observed that choosing a value as small as 4 did not effect the overall phase diagram). We used finite and infinite Density Matrix Renormalization Group (DMRG and iDMRG respectively) to study various features of our model. All numerical calculations were performed using the TeNPY library \cite{hauschild2018efficient}. 

The phase diagram obtained using DMRG is shown in Fig.~\ref{Fig:DMRG} a) and the relevant correlation function plot in the critical phase is shown in Fig.~\ref{Fig:DMRG} b). Overall, the numerically observed phase diagram corroborates the analytical results: The Berezinskii-Kosterlitz-Thouless transition line out of the superconducting phase (red) as obtained from Eq.~\eqref{eq:SCAction} (corresponding to $K_{\rm{SC}} = 1$) captures well the boundary of the area where the central charge is $c = 1$. The correlators inside the regime with $c = 1$ are consistent with off-diagonal long-range order of a singlet superconductor, Fig.~\ref{Fig:DMRG} b).
As expected, for $\Delta < \sqrt{E_C t'} \approx 0.1 \sqrt{t^2 + t'^2}$, the wide superconducting phase narrows to a sharp line located at $t = t'$, cf. Fig.~\ref{Fig:Model} c) and \ref{Fig:DMRG} a). Finally, the nature of the insulating ($c = 0$) phases is verified using iDMRG, by means of which we obtained the entanglement spectrum of the system, see Fig.~\ref{Fig:DMRG}(c).
We observe an even degeneracy of all levels throughout the entanglement spectrum in the regime corresponding to the topological insulator in Fig.~\ref{Fig:Model} c), but not in the topologically trivial regime. Since even degeneracies are a signature of SPT order ~\cite{PollmannOshikawa2010}, this supports our analytical result of QRSPT order in our model. We also observed a similar entanglement spectrum for $N_{g} =2$ as well. We also observe~\cite{SuppMat} edge states by measuring local charge and spin expectation values of the DMRG ground state and an edge phase transition analogous to Fig.~\ref{Fig: Analytical results} a).

Contrary to the analytical expectation, the numerically obtained central charge continuously reaches three at $t = t'$ as $\Delta$ is decreased. We attribute this observation to numerical limitations. Non-integer central charge values are not to be expected in the present context. A discussion of numerical limitations in terms of the energy variance of the obtained groundstates in the regime of large $c$ as well as the error bars of Fig.~\ref{Fig:DMRG} c) are relegated to Ref.~\cite{SuppMat}. 

Although the numerical observation of both charge and spin edge modes is consistent with $\mathit{C}$ being the protecting symmetry, we observe that breaking only the $\mathit{C}$ symmetry (by adding a symmetry breaking perturbation) does not lift the degeneracies in the entanglement spectrum. This possibly implies the existence of other symmetries protecting the SPT phase and/or it is even possible that the symmetry arguments are modified due to the presence of bosonic degrees of freedom in our model since the existing theoretical arguments are valid, for $C$ being the protecting symmetry, only for non interacting/interacting fermionic systems ~\cite{Zirnbauer2021}. However the degeneracies are lifted in the presence of a term that breaks total charge ($U(1)$) symmetry implying that either $U(1)$ has a non trivial representation at the edge -- which is not expected on general theoretical grounds~\cite{ChenWen2011}. Further details regarding the symmetry breaking perturbations can be found in ~\cite{SuppMat} and we leave a detailed numerical/analytical study to characterize relevant symmetries for future works.

\PRLSection{Conclusion:} In summary, we have introduced the concept of quantum restoration of symmetry protected topological phases, i.e. topological systems in which underlying protecting symmetry is broken at each instance of time, but restored upon time average. 
To illustrate this concept, we carefully studied an interacting one-dimensional model corresponding to a spinful Su-Schrieffer-Heeger model with fluctuating superconductivity. Using combined analytical and numerical methods, we demonstrate that the SPT features are restored. It is worthwhile to highlight a phase diagram that is distinct from and arguably richer than purely fermionic interacting SSH-chains~\cite{BenthienGrage2006,ManmanaGurarie2012,SirkerSedlmayr2014,YahyaviHetenyi2018,Nersesyan2020,MatveevaCarr2024,PadhanMishra2024}, both in terms of phases and phase transitions in the bulk and of those at the boundary. 
While it is generally expected that insulating QRSPTs are related to topological zeros in the fermionic Green's function ~\cite{Gurarie2011,WagnerSangiovanni2023,SoldiniNeupert2023,ManningBradlyn2023,SettySi2024,BollmannKoenig2023}, we leave this as an open question for future studies along with, first, which QRSPT phases can be expected based on a given underlying free fermion SPT and mean-field SSB, second, whether there are universal patterns in the corresponding phase diagram between QRSPT and trivial phase, and third, a careful analysis in higher dimensions.

\PRLSection{Methods:}

\textit{Symmetries.}
The model in Eq.~\eqref{Eq: Hamiltonian for the total model.} has $U(1)$ symmetry (corresponding to a conserved total charge $\sum_{X} 2\hat{N}_{X} + \hat{n}_{X}$) and combined sublattice and particle-hole symmetry (in the second quantization language) $\mathit{C}$ given by:
\begin{gather}
    \begin{aligned}
        C d^{\dagger}_{X,A,\sigma}C^{-1} &= d_{X,A,\sigma}, \hspace{4 pt} C d_{X,A,\sigma} C^{-1} = d^{\dagger}_{X,A,\sigma}, \\
        C d^{\dagger}_{X,B,\sigma}C^{-1} &= -d_{X,B,\sigma}, \hspace{4 pt} C d_{X,B,\sigma} C^{-1} = -d^{\dagger}_{X,B,\sigma}, \\
        C \hat{\phi}_{X} C^{-1} &= \hat{\phi}_{X} + \pi, \hspace{12 pt} CiC^{-1} = -i,
    \end{aligned}
\end{gather}
where $C^{2} = \mathcal{I}$. {As compared to \cite{VerresenPollmann2017,Zirnbauer2021} , note that we have extended $\mathit{C}$ to bosons such that it reverses the charge of the bosons and preserves the commutation relations}.

\textit{Mean-field Hamiltonian.} The mean-field Bloch Hamiltonian takes the form $H = \int \frac{dp}{2\pi} \Psi^\dagger(p) H(p) \Psi(p)$ where $\Psi(p) = \left ( d_{A,\uparrow}(p), d_{B, \uparrow}(p), d^\dagger_{A, \downarrow} (-p), d^\dagger_{B, \downarrow}(-p) \right)^T$. Clearly, in the absence of SSB, i.e. for $\Delta = 0, H(p)$ is block diagonal as illustrated schematically in Fig.~\ref{Fig:QRSPT}. It is convenient to choose a different basis in which 
 \begin{subequations}
 \begin{align}
     H(p) & \rightarrow \left ( \begin{array}{cc}
         0 & D(p) \\
         D^\dagger(p) &0
    \end{array} \right ), \label{eq:HBdG} \\
    D(p) & = \left (\begin{array}{cc}
       - (t + t' e^{i p})  & i \Delta  \\
       - i \Delta & (t + t' e^{-i p})   
    \end{array}\right ).
\end{align}
\end{subequations}
The winding represented in Fig.~\ref{Fig:QRSPT} a), b) correspond to the parametric plot of eigenvalues of $D(p)$ at $t = 0.1 t'$ and $\Delta = 0$ for a) and $\Delta = 0.1 t'$ for panel b). 

\textit{Parameters of the continuum field theory.} The parameters in Eq.~\eqref{Eq: Bosonized Equation} are defined as follows in terms of the model parameters in Eq.~\eqref{Eq: Hamiltonian for the total model.}:
\begin{gather}
    \begin{aligned}
        K_{s} &= \frac{1}{1 + \frac{\Delta^{2}}{8E_{\rm{C}}t'\pi} }, &
        K_{\rho} &= \frac{1}{\sqrt{(1 - \frac{\Delta^{2}}{8E_{\rm{C}}t'\pi})(1- \frac{\Delta^{2}}{8E_{\rm{C}}t'\pi} + \frac{E_{\rm{C}}}{4t'\pi})}} , \\
        u_{s} &= t' a, &  u_{\rho} &= u_s\sqrt{\frac{1- \frac{\Delta^{2}}{8E_{\rm{C}}t'\pi} + \frac{E_{\rm{C}}}{4t'\pi}}{1 - \frac{\Delta^{2}}{8E_{\rm{C}}t'\pi}}},\\
        G_{s} &= \frac{\Delta^{2}}{8\pi^{2}E_{\rm{C}}a}, &
        M &= \frac{2(t^{\prime}-t)}{\pi a}.
    \end{aligned}
\label{SM:Eq: Definitions of Parameters in the bosoniized equation}
\end{gather}

Note that we these expressions are valid close to $t = t'$ and we generally assume $E_{\rm{C}} \ll \sqrt{t^{2}+t^{\prime 2}}$. 

\textit{Luttinger parameters of the superconductor.} 
To integrate out fermions, we expand the transformed Hamiltonian, obtained after the unitary transformation, in terms of $(\hat{\phi}_{X+1} - \hat{\phi}_{X})$ up to second order and also rewrite it in the action formalism. The corresponding action is given by:
\begin{gather}
    \begin{aligned}
        S &= S_{0}(\phi) + S_{0}(d,\Bar{d}) + \Delta S_{1}(\phi,d,\Bar{d}) \\
        &+ \Delta S_{2} (\phi,d,\Bar{d}) ,
    \end{aligned}
\end{gather}
where $d/\Bar{d}$ represent Grassmann variables and $\phi$ represents the superconducting phase. The effective bosonic action is given by:
\begin{gather}
    \begin{aligned}
        S_{\rm{eff}}(\phi) &= S_{0}(\phi) + \langle \Delta S_{1} (\Bar{d},d,\phi) \rangle \\
        &+ \langle \Delta S_{2} (\Bar{d},d,\phi) \rangle \\
        & -\frac{1}{2} (\langle \Delta S_{1}^{2} (\Bar{d},d,\phi)\rangle - \langle \Delta S_{1} (\Bar{d},d,\phi)\rangle^{2}),
    \end{aligned}
\end{gather}
where,
\begin{equation}
    \langle . \rangle = \frac{1}{Z_{0}(\Bar{d},d,\phi)} \int \mathcal{D} \Bar{d} \mathcal{D} d \hspace{2 pt}. \hspace{2 pt} e^{-S_{0}(\Bar{d},d)}.
\end{equation}
The parameters entering the effective superconducting theory in $S_{\rm{eff}}(\phi)$ given in Eq.~\eqref{eq:SCAction} are related to the microscopic parameters as follows:
\begin{gather}
    \begin{aligned}
        K_{\rm{SC}} &= 2\sqrt{\frac{4\pi^{2}E_{\rm{C}}a}{vI}},\\
        u_{\rm{SC}} &= 4\sqrt{\frac{vIE_{\rm{C}}a}{4\pi^{2}}},\\
        v &= t^{\prime}a,\\
        I &= \frac{2 \pi \delta ^2  \cos (\theta ) \rm{E}\left(-\frac{2 \sin(2 \theta)}{\delta ^2+ 1 -\sin (2\theta)}\right)}{\sqrt{\delta ^2+1 - 2\sin(2\theta)} \left(\delta ^2+1 + 2\sin(\theta)\right)},
    \end{aligned}
\end{gather}
where $\theta = \arctan (\frac{t}{t^{\prime}})$ and $\delta = \frac{\Delta}{\sqrt{t^{2} + t^{\prime2}}}$. $K_{\rm{SC}}$ represents inverse superconducting stiffness. Further details are provided in~\cite{SuppMat}.

\PRLSection{Data Availability}
The numerical data that support the findings of this study are available from Zenodo repository 10.5281/zenodo.11243225 \cite{Zenodo} .\\
\PRLSection{Code Availability}
The code used for numerical simulations is available from Zenodo repository 10.5281/zenodo.11243225 \cite{Zenodo}.

\bibliography{QRSPT}

\PRLSection{Acknowledgements:}
It is a pleasure to thank Sam Carr, Frank Pollmann, Ruben Verresen, Johannes Hauschild, Martin R. Zirnbauer, Yuval Oreg for useful input and to thank Thomas K\"ohler for collaboration on a related project. DT thanks Kirill Parshukov, Nikolaos Parthenios, and Raffaele Mazzilli for useful discussions.  Support for this research was provided by the Office of the Vice Chancellor for Research and Graduate Education at the University of Wisconsin–Madison with funding from the Wisconsin Alumni Research Foundation. This research was supported in part by grant NSF PHY-2309135 to the Kavli Institute for Theoretical Physics (KITP). DT thanks Max Planck Institute for Solid State Research for the IMPRS fellowship. SP acknowledge support from the Munich Center for Quantum Science and Technology. EJK acknowledges hospitality by the KITP.

\PRLSection{Author Contributions.} EJK conceived and supervised the project. DT performed all calculations under the guidance of SB, SP, EJK. All co-authors analyzed the results and wrote the manuscript.

\PRLSection{Competing Interests.}

The authors declare no competing interests.

\clearpage

\setcounter{equation}{0}
\setcounter{figure}{0}
\setcounter{section}{0}
\setcounter{table}{0}
\setcounter{page}{1}
\makeatletter
\renewcommand{\theequation}{S\arabic{equation}}
\renewcommand{\thesection}{S\arabic{section}}
\renewcommand{\thefigure}{S\arabic{figure}}
\renewcommand{\thepage}{S\arabic{page}}

\begin{widetext}
\begin{center}
Supplementary materials on \\
\textbf{"Quantum Restored Symmetry Protected Topological Phases"}\\
{Dhruv Tiwari$^{1,2}$,
{Steffen Bollmann $^1$,}
{Sebastian Paeckel $^2$},
{Elio J. K\"onig}$^{1,4}$}\\
{%
$^1$ Max-Planck Institute for Solid State Research, 70569 Stuttgart, Germany
}\\%
{%
$^2$Max Planck Institute for the Physics of Complex Systems, 01187 Dresden, Germany
}\\%
{%
$^3$Department of Physics, Arnold Sommerfeld Center for Theoretical Physics (ASC), Munich Center for Quantum Science and Technology (MCQST), Ludwig-Maximilians-Universit\"at M\"unchen, 80333 M\"unchen, Germany
}\\%
{%
$^4$Department of Physics, University of Wisconsin-Madison, Madison, Wisconsin 53706, USA
}
\end{center}
\end{widetext}

This supplementary material is structured as follows, Sec:\ref{SM:Sec:Symmetries and their action} contains details about some symmetries of the system, in particular particle-hole transformation $\mathit{C}$. In Sec:\ref{SM:Sec:Perturbative calculation}, we detail the perturbative calculations and show that the symmetry $\mathit{C}$ has a non trivial edge representation in the phase exhibiting gapless edge modes. In Sec:\ref{SM:Sec:Bosonization}, we detail the procedure of bosonization and semiclassics. In Sec:\ref{SM:Sec:Integrating out fermions}, we detail the action of the unitary transformation and the procedure of integrating out fermions. In Sec:\ref{Sec:DMRG and Data extraction}, we present additional numerical data regarding error analysis, numerical data on edge transition and study of the entanglement spectrum in presence of symmetry breaking perturbations.

\section{Symmetries and their action} 
\label{SM:Sec:Symmetries and their action}

 As mentioned in the main text, the model is invariant upon an even shift of $N_{g}$. This can be seen simply by the fact that an even shift in $N_{g}$ can be represented by $N_{g} \rightarrow N_{g} + 2\mathbb{Z}$ and can be followed by a redefinition of the operator $\hat{N}_{X} \rightarrow \hat{N}_{X} + \mathbb{Z}$. This does not lead to a change in the physics of the model because the defining commutation relations of the model are left unchanged. This invariance of the model under an even shift of $N_{g}$ plays a crucial role in showing that the model is invariant under $\mathit{C}$ for all even values of $N_{g}$. This can be understood by the transformation of $\hat{N}_{X}$ and $\hat{n}_{X}$ under $\mathit{C}$:
\begin{gather}
    \begin{aligned}
        \mathit{C} \hat{N}_{X} \mathit{C}^{-1} &= -\hat{N}_{X}, \\
        \mathit{C}\hat{n}_{X} \mathit{C}^{-1} &= 4\mathbb{I} - \hat{n}_{X}.
    \end{aligned}
\end{gather}\\
Thus $C\hat{H}C^{-1}$ corresponds to a theory that is described by a new $N^{\prime}_{g} = 4 - N_{g}$, corresponding to an even shift. Note that $\mathit{C}$ is an exact symmetry of the model for $N_{g} = 2$. 

\section{Perturbation about the dimerized limits}
\label{SM:Sec:Perturbative calculation}

The energy of the unperturbed state in Eq.~\eqref{Eq: Groundstate to perturb about for t=0} is given by:
\begin{equation}
    E_{\psi} = -t^{\prime}N.
\end{equation}
After a first-order perturbation analysis by introducing $\Delta$ and $E_{\rm{C}}$ as perturbations, the correction to the groundstate is given by:
\begin{equation}
    E^{1}_{\psi} = (-t^{\prime} + E_{\rm{C}})N.
    \label{Eq: Correction to the groundstate energy for t=0}
\end{equation}
Note that the first order correction to the groundstate energy implies that $\ket{\psi}$ is the groundstate of the total Hamiltonian only if $t^{\prime} > E_{\rm{C}}$ or generally $\sqrt{t^{2} + t^{\prime 2}} > E_{\rm{C}}$ since there is a competing state given for integer values of $N_{g}$ given by:
\begin{equation}
    \ket{\phi} = \prod_{X=1}^{N} \ket{N_{X}= \frac{N_{g}}{2}} \ket{n_{X} = 0},
\end{equation}
which has zero energy corresponding to the total Hamiltonian given in Eq~\eqref{Eq: Hamiltonian for the total model.}.\\
To study signatures of symmetry fractionalization, we focus on the left edge in the regime $t = 0$ (since we do not obtain any edge states in the other regime corresponding to $t^{\prime} = 0$) and study the case of degenerate spin and charge edge modes separately. For $\Delta < E_{\rm{C}}$, the operators acting on the lowest energy Hilbert space of edge (spanned by $\ket{\uparrow}$ and $\ket{\downarrow}$) are given by:
\begin{gather}
    \begin{aligned}
        \vec{S} = \frac{1}{2} d^{\dagger}_{L}\vec{\sigma}d_{L},
    \end{aligned}
\end{gather}
where $\vec{\sigma} = (\sigma_{x},\sigma_{y},\sigma_{z})$ ($\sigma$ represent the Pauli matrices) and similarly for $\vec{S}$. $d^{(\dagger)}_{L}$ denote the fermionic operators on the left edge. The symmetry $\mathit{C}$ can be decomposed in the low energy sector as:
\begin{equation}
    \mathit{C} = U_{L}U_{R}\Tilde{K},
    \label{SM:Eq:frac of symmetry C}
\end{equation}
where $U_{L} = id^{\dagger}_{L}\sigma_{y}d_{L}$ and $U_{R} = id^{\dagger}_{R}\sigma_{y}d_{R}$ ($d^{(\dagger)}_{R}$ denotes fermionic  operators on the right edge) and $\Tilde{K}$ is defined in terms of the fermionic operators on the edges:
\begin{gather}
    \begin{aligned}
        \Tilde{K}d^{(\dagger)}_{L/R}\Tilde{K} = d^{(\dagger)}_{L/R}.
    \end{aligned}
\end{gather}
Note that $\mathit{C}^{2} = \mathbb{I}$ and it can be seen that this still holds when we write $\mathit{C}$ in the form written in Eq.~\eqref{SM:Eq:frac of symmetry C}. The fractionalization of an anti-unitary symmetry is indicated using $\Bar{U}_{L/R}U_{L/R} = \pm 1$ where $\Bar{U}_{L/R}$ denotes $CU_{L/R}C$. A non-trivial edge state (topological phase) is present when the algebra of the edge symmetries ($U_{L/R}$) is different from that of the bulk symmetry ($C$). It can be seen that in the topological phase for $\Delta < E_{\rm{C}}$, we indeed have a fractionalization of symmetry indicated by:
\begin{equation}
    \Bar{U}_{L/R}U_{L/R} = -1.
\end{equation}
Similarly, for $\Delta > E_{\rm{C}}$, the degenerate charge edge modes are also protected by symmetry $\mathit{C}$. To see this, let us denote the charge edge states on the left edge in a concise form:
\begin{gather}
    \begin{aligned}
        \ket{2_{L}} &= \frac{e^{i\hat{\phi}_{L}}\ket{0}_{L} - i\ket{\uparrow \downarrow}}{\sqrt{2}},\\
        \ket{\Omega_{L}} &= \frac{\ket{0}_{L} - ie^{-\hat{\phi}_{L}}\ket{\uparrow \downarrow}}{\sqrt{2}}, 
    \end{aligned}
\label{SM:Eq:representation of charge edge modes in a concise form}
\end{gather}
where $\ket{0}_{L}$ represents the vacuum on the left edge comprising of the first bosonic and fermionic site and the subscript $L$ denotes the operator on the left edge. The operators acting on the lowest energy Hilbert space are given by:
\begin{gather}
    \begin{aligned}
        \sigma^{\prime}_{z,L} &= \ket{\Omega_{L}}\bra{\Omega_{L}} - \ket{2_{L}}\bra{2_{L}}, \\
        \sigma^{\prime}_{x,L} &= -i\ket{\Omega_{L}}\bra{2_{L}} + i\ket{2_{L}}\bra{\Omega_{L}}, \\
        \sigma^{\prime}_{y,L} &= \ket{\Omega_{L}}\bra{2_{L}} + \ket{2_{L}}\bra{\Omega_{L}}.
    \end{aligned}
\label{SM:Eq:representation of operators for low energy h space}
\end{gather}
Similarly, one can define the corresponding operators for the right edge. As before, the symmetry can be decomposed as:
\begin{equation}
    \mathit{C} = U^{\prime}_{L}U^{\prime}_{R}\Tilde{K}^{\prime},
\end{equation}
where $\Tilde{K}^{\prime}$ is defined as $\Tilde{K}^{\prime}(\sigma^{\prime}_{x,L/R},\sigma^{\prime}_{y,L/R},\sigma^{\prime}_{z,L/R})\Tilde{K}^{\prime} = (\sigma^{\prime}_{x,L/R},-\sigma^{\prime}_{y,L/R},\sigma^{\prime}_{z,L/R})$. Using this definition, the edge symmetry representation of $\mathit{C}$ is given by $U^{\prime}_{L/R}= i\sigma^{\prime}_{y,L/R}$. It can be seen that $\Bar{U}^{\prime}_{L/R}U_{L/R} = -1$ implying that the charge edge modes are protected by anti-unitary symmetry $C$ as well. 

\section{Field theory near the free fermion critical point}
\label{SM:Sec:Bosonization}

In this section, we consider the limit in which the bosonic gap set by the charging energy is large as compared to other energy scales and we can thus integrate out the bosons. The effective Hamiltonian describing the low energy theory of fermions is given by:
\begin{gather}
    \begin{aligned}
        \hat{H}_{\textrm{eff}} &= \sum_{X} E_{\rm{C}}( :\hat{n}_{X}:)^{2} - t\sum_{X,\sigma}(d^{\dagger}_{X,\sigma,A}d_{X,\sigma,B}+ \mathrm{H.c.})\\
        &-t^{\prime}\sum_{X,\sigma}(d^{\dagger}_{X+1,\sigma,A}d_{X,\sigma,B} + \mathrm{H.c.})\\
        &-\frac{\Delta^{2}}{2E_{\rm{C}}}\sum_{X,j,j^{\prime}}(d^{\dagger}_{X,j}[\sigma_{y}]d^{\dagger}_{X,j}d_{X,j^{\prime}}[\sigma_{y}]d_{X,j^{\prime}} + \mathrm{H.c.}),
    \end{aligned}
\label{Eq: Effective Hamiltonian for Integrating out bosons.}
\end{gather}
where $\hat{n}_{X} = \sum_{j,\sigma} \hat{n}_{X,j,\sigma}$. Note that we still have to do a summation over $\sigma$ indices in the $\Delta$ term and they have been dropped for brevity. The $::$ in Eq.~\eqref{Eq: Effective Hamiltonian for Integrating out bosons.} refer to normal ordering with respect to the non-interacting fermionic groundstate (corresponds to non zero values of $t,t^{\prime}$ and $\Delta = E_{C} = 0$).

\subsection{Bosonization}

Before performing bosonization, it is necessary to obtain the effective continuum Hamiltonian in terms of right and left movers:
\begin{small}
\begin{gather}
    \begin{aligned}
    S_{\textrm{eff}} &= \int d \tau d x \Bar{L}_{\sigma} [\partial_{\tau} + i v \partial_x] L_{\sigma} + \Bar{R}_{\sigma}[\partial_{\tau} - i v \partial_x]R_{\sigma} \\
    &+ (t^{\prime} - t)(-i \Bar{L}_{\sigma}R_{\sigma} + \mathrm{H.c.}) + \frac{E_{\rm{C}} a}{16} :(\Bar{L}_{\sigma}L_{\sigma} + \Bar{R}_{\sigma}R_{\sigma}):^{2} \nonumber \\
    &+ \frac{\Delta^{2} a }{4 E_{C}} (-\Bar{L}_{\uparrow}L_{\uparrow}\Bar{R}_{\downarrow}R_{\downarrow} - \Bar{L}_{\downarrow}L_{\downarrow}\Bar{R}_{\uparrow}R_{\uparrow} )\\
    &+  \frac{\Delta^{2} a }{4 E_{C}} (\Bar{L}_{\uparrow}L_{\downarrow} \Bar{R}_{\downarrow}R_{\uparrow}+ \Bar{L}_{\downarrow}L_{\uparrow}\Bar{R}_{\uparrow}R_{\downarrow} ),
    \end{aligned}
\label{SM:Eq:action for effective fermionic hamiltonian} 
\end{gather}
\end{small}
where $a$ defines the size of the unit cell, $v = t^{\prime}a$, lowercase $x$ denotes continuous position variable and $\tau$ denotes imaginary time. $L_{\sigma}/R_{\sigma}$ are the Grassmann variables representing the left and right movers with spin $\sigma$. Note that all the fermionic fields in the above equation have $x$ and $\tau$ dependence.
The bosonized action, Eq.~\eqref{Eq: Bosonized Equation}in the main text, with parameters summarized in the methods section was obtained using the following identities:
\begin{gather}
    \begin{aligned}
        \Bar{R}_{\sigma} &= \frac{\xi}{\sqrt{2\pi \alpha}} e^{i\sqrt{2}\phi_{\sigma}},
        \Bar{L}_{\sigma} = \frac{\Bar{\xi}}{\sqrt{2\pi\alpha}} e^{-i \sqrt{2}\Bar{\phi}_{\sigma}},\\
        \Bar{L}_{\sigma}L_{\sigma} &= \frac{-i}{\sqrt{2}\pi}\partial \Phi_{\sigma},
        \Bar{R}_{\sigma}R_{\sigma} = \frac{-i}{\sqrt{2}\pi} \Bar{\partial} \Phi_{\sigma},\\
        \Phi_{\sigma} &= \phi_{\sigma} + \Bar{\phi}_{\sigma},\\
        \partial &= \frac{1}{2}(\frac{\partial_{\tau}}{v} - i\partial_{x}),
        \Bar{\partial} = \frac{1}{2} (\frac{\partial_{\tau}}{v} + i\partial_{x}),\\
        \Phi_{\rho} &= \frac{\Phi_{\uparrow} + \Phi_{\downarrow}}{\sqrt{2}},
        \Phi_{s} = \frac{\Phi_{\uparrow} - \Phi_{\downarrow}}{\sqrt{2}},
    \end{aligned}
    \label{Eq: Bosonization identities}
\end{gather}
where $\alpha$ is the cutoff of the theory. We fix $\alpha = a$, the size of unit cell in our model and $\xi (\Bar{\xi})$ are Klein factors to ensure proper anti-commutation relations. 

\subsection{Semiclassical analysis of edge states}

\begin{figure}
    \centering
    \includegraphics[scale = 0.45]{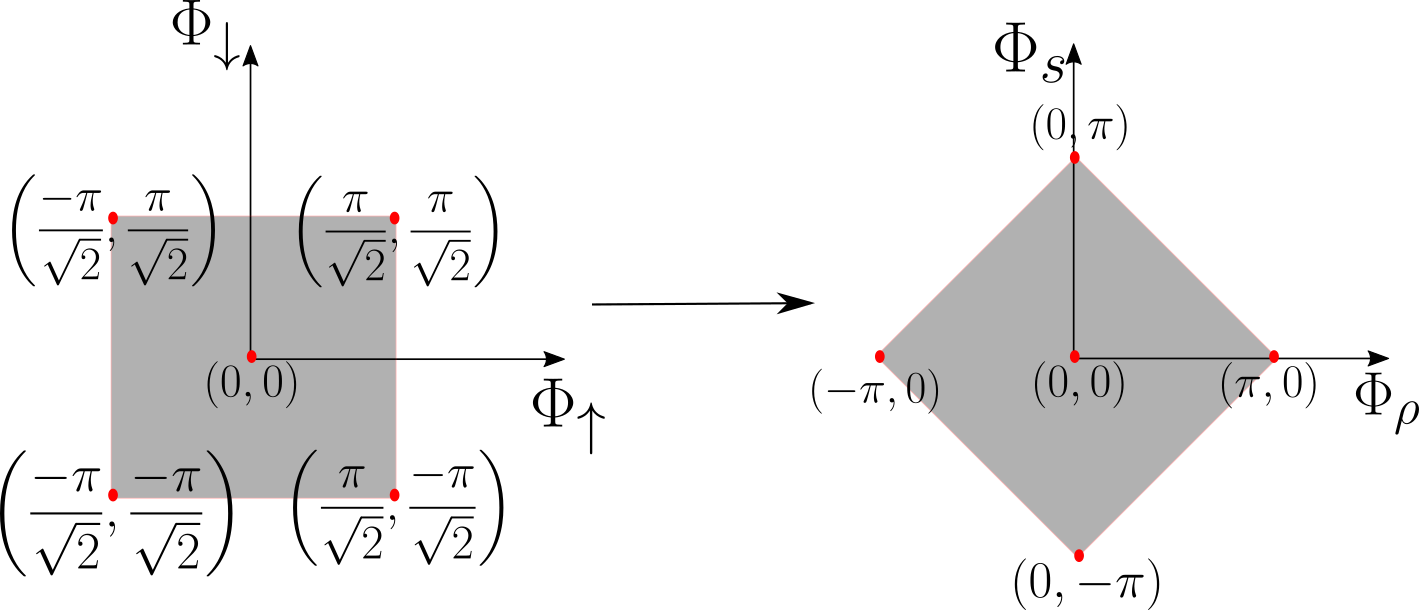}
    \caption{Shaded region represents a segment with unique value of the bosonic variables. The translation from the shaded region for the bosonic variables $(\Phi_{\uparrow},\Phi_{\downarrow})$ to $(\Phi_{s},\Phi_{\rho})$ is done using the transformation identity in Eq.~\eqref{SM:Eq: Definitions of Parameters in the bosoniized equation}.}
    \label{SM:Fig: Semiclassiscs pbc}
\end{figure}

To perform the semiclassical analysis, it is essential to understand the groundstate of the action in Eq.~\eqref{Eq: Bosonized Equation} in terms of bosonic variables for two different signs of $M$ and $M \neq 0$. As can be seen in Eq.~\eqref{Eq: Bosonization identities}, the bosonic variables $\Phi_{\sigma}$ have a periodicity of $\sqrt{2}\pi$. This periodicity can be translated in terms of $\Phi_{\rho}$ and $\Phi_{s}$, as shown in Fig.~\ref{SM:Fig: Semiclassiscs pbc}. 
If we restrict ourselves to the shaded region in Fig.~\ref{SM:Fig: Semiclassiscs pbc}, then for $M<0$ and $G_{s}>0$, the value of the bosonic variables for the groundstate is $(\Phi_{\rho},\Phi_{s}) = (0,0)$. For $M>0$, there are four possible values given by $(\Phi_{\rho},\Phi_{s}) = (0,\pi), (\pi,0), (-\pi,0),(0,-\pi)$. Note that these configurations are sufficient to study all possible charge and spin excitations that occur separately since any other configuration can be mapped to these by invoking the periodicity of the bosonic variables. Also, as mentioned in the main text, these configurations lead to the same groundstate energy in the bulk.\\
We calculated the energy of charge excitation (for fixed $\Phi_{s,i} = 0$) and spin excitations (for fixed $\Phi_{\rho,i} = 0$) at the interface of the phases corresponding to $M<0$ (for $x<0$ [blue region] in the inset of Fig.~\ref{Fig: Analytical results} (b)) for arbitrary values of $G_{s}$ and $M=\infty$ (for $x>0$ [yellow region] in the inset of Fig.~\ref{Fig: Analytical results}(b)). The free energy is minimized by spatial fluctuations of the field variables alone (temporal fluctuations are costly due to the derivative term). Correspondingly, this problem is analogous to a half-instanton solution in the potentials
\begin{gather}
    \begin{aligned}
    V(\Phi_{\rho}) &= -M\cos(\Phi_{\rho}) - G_{s},\\
    V(\Phi_{s}) &= -M\cos(\Phi_{s}) - G_{s}\cos(2\Phi_{s}).
    \end{aligned}
\end{gather}

The corresponding excitation energy of the edge excitations is given by (setting $\Phi_{\rho,f} = \pi$ and $\Phi_{s,f} = \pi$):
\begin{gather}
    \begin{aligned}
    E_{\rho} (\Phi_{\rho,f}) &= \sqrt{\frac{u_{\rho}}{K_{\rho}}} \int_{0}^{\Phi_{\rho,f}} d\Phi_{\rho}\sqrt{V(\Phi_{\rho}) - V(\Phi_{\rho,f})},  \\
    & = 2\sqrt{2\vert M \vert \frac{u_\rho}{K_\rho}}, \\
    E_{s} (\Phi_{s,f}) &= \sqrt{\frac{u_{s}}{K_{s}}} \int_{0}^{\Phi_{s,f}} d\Phi_{s}\sqrt{V(\Phi_{s}) - V(\Phi_{s,f})}, \\
    & = \sqrt{2\vert M\vert \frac{u_s}{K_s}} \left ( \sqrt{1 + 4 \frac{G_s}{\vert M\vert}} + \frac{\text{arcsinh}(2 \sqrt{\frac{G_s}{\vert M\vert}})}{2 \sqrt{\frac{G_s}{\vert M\vert}})} \right),
    \end{aligned}
\label{SM:Eq:Energy of spin and charge excitations}
\end{gather}
where we found these expressions using the standard methodology for obtaining the instanton action of a one-dimensional problem in an arbitrary potential. 

\begin{figure}
    \centering
    \includegraphics[scale = 0.05]{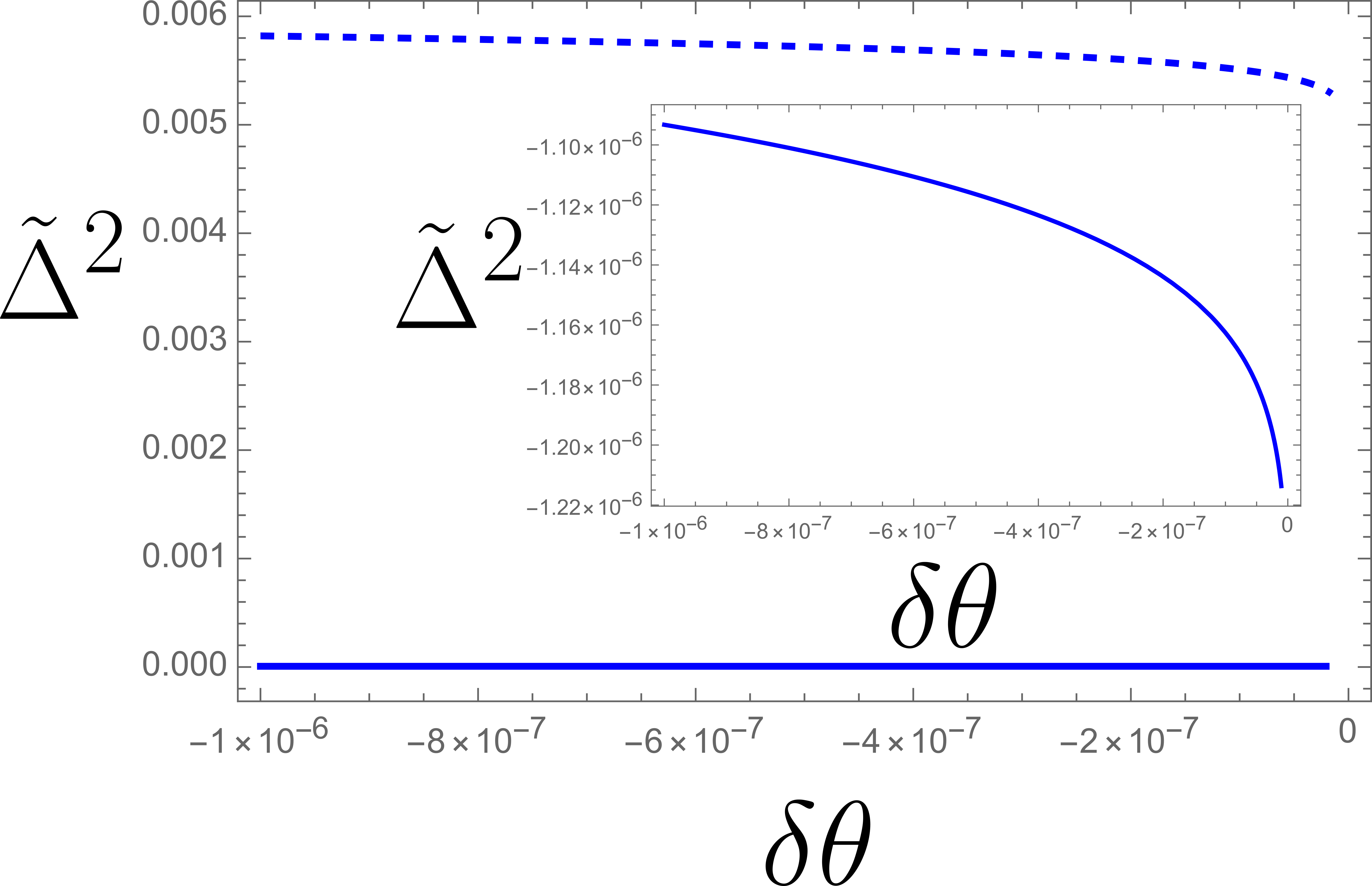}
    \caption{Plot of $\frac{L_{M}}{L_{G_{s}}} = 1$ (dashed line) and edge transition curve corresponding to $\frac{E_{s}}{E_{\rho}} = 1$ (solid line and inset).}
    \label{Fig:SM:Semiclassical edge transition}
\end{figure}
We next wish to determine which of the two type of kink solutions has lower energy. We find
\begin{align}
    \frac{E_s}{E_\rho} & = \sqrt{\frac{u_s K_\rho}{u_\rho K_s}} \frac{\sqrt{1 + 4 \frac{G_s}{\vert M\vert}} + \frac{\text{arcsinh}(2 \sqrt{\frac{G_s}{\vert M\vert}})}{2 \sqrt{\frac{G_s}{\vert M\vert}})} }{2}.
    \label{SM:Eq: ratio of energy for charge and spin edge modes}
\end{align}
The first factor is
\begin{equation}
    \sqrt{\frac{u_{s} K_{\rho}}{u_\rho K_{s}}}  = \sqrt{\frac{1 + \frac{\Delta^2}{8 E_{\rm{C}} \pi t'}}{1 - \frac{\Delta^2}{8 E_{\rm{C}} \pi t'} + \frac{E_{\rm{C}}}{4t^{\prime}\pi}}}, 
\end{equation}
as it is not normalized within the present scheme of leading order RG. For the second term, we should use renormalized values of the parameters $M$ and $G_{s}$ in order to ensure that the localization length of the edge states resembles the non interacting result once we set $\Delta=E_{\rm{C}} = 0$. An estimate of the localization length of the edge states in the charge sector can be obtained using the Euler-Lagrange equations for $\Phi_{\rho}$  field and is given by:

\begin{align}
    \xi =  \sqrt{\frac{u_{\rho}}{K_{\rho}\Bar{M}}},
\end{align}
where $\Bar{M} = \frac{u_{s}}{\Tilde{a}^{2}}M_{\text{dimensionless}}$ and $\Tilde{a}$ represents a renormalized length scale and $u_{s}$ is given in Eq.~\eqref{SM:Eq: Definitions of Parameters in the bosoniized equation}. Note that the bare value of $\Bar{M}$ is given by $M$ in Eq.~\eqref{SM:Eq: Definitions of Parameters in the bosoniized equation} and $\Tilde{a} = a$ where $a$ represent the lattice spacing. For renormalized parameters, we should use the flow equations given in Eq.~\eqref{Eq: RG flow equations} to obtain the renormalized value of $M_{\text{dimensionless}}$. The value of $\Tilde{a}$ is given by lengthscale at which the RG flow stops. To determine the RG lengthscale, we should determine as to which term, $M$ or $G_{s}$ diverges first within the regime of applicability of the effective theory. This can be achieved by comparing the RG lengthscale of both $M$ and $G_{s}$ term which are given below upto leading orders:

\begin{align}
    L_{M} &= a\bigg(\frac{\Tilde{E}_{\rm{C}}\pi}{2\sqrt{2}|\delta\theta|}\bigg), \\
    L_{G_{s}} &=  a \bigg(\frac{8\pi^{2}\Tilde{E}_{\rm{C}}^{2}}{\Tilde{\Delta}^{2}}\bigg)^{\frac{4\sqrt{2}\Tilde{E}_{\rm{C}}\pi}{\Tilde{\Delta}^{2}}},
\end{align}

where $\Tilde{E}_{\rm{C}}$ and $\Tilde{\Delta}$ respectively, are normalized with respect to $\sqrt{t^{2} + t^{\prime2}}$ and $\theta = \arctan(\frac{t}{t^{\prime}}) $ and $\delta\theta \ll 1$.\\
The curve corresponding to $\frac{L_{M}}{L_{G_{s}}} = 1$ close to the phase transition (since $\delta \theta <<1$) is shown in Fig.~\ref{Fig:SM:Semiclassical edge transition}. Thus within the region where $M$ diverges first, RG lengthscale is given by $L_{M}$ and thus the localization length of the edge states is proportional to $\xi \approx \frac{1}{|\delta \theta|}$ which gives the non interacting result once we set $E_{\rm{C}} = \Delta = 0$. It can be similarly shown that renormalized values should be used for $G_{s}$ as well.

Eq.~\eqref{SM:Eq: ratio of energy for charge and spin edge modes} can be approximated upto leading order as given below assuming $\Tilde{E}_{\rm{C}},\Tilde{\Delta},|\delta\theta| \ll 1, \Tilde{\Delta} \ll \Tilde{E}_{\rm{C}}$:
\begin{small}
\begin{align}
   \frac{E_{s}}{E_{\rho}} = \bigg(1 - \frac{\Tilde{E}_{\rm{C}}}{\sqrt{32}\pi}\bigg)\bigg(1 + \frac{\Tilde{\Delta}^{2}}{12\pi^{2}\Tilde{E}_{C}^{2}}\bigg(\frac{\Tilde{E}_{\rm{C}}\pi}{2\sqrt{2}|\delta \theta|}\bigg)^{\frac{\Tilde{\Delta}^{2}}{4\sqrt{2}\Tilde{E}_{\rm{C}}\pi}}\bigg),
   \label{Eq:SM: Approximation of edge transition equation}
\end{align}
\end{small}
where we expanded the second term upto first orders in $\frac{G_{s}}{M}$ and used the renormalized value of $G_{s}$. For $\Tilde{E}_{C} = 0.01$, the curve corresponding to $\frac{E_{s}}{E_{\rho}} = 1$ can be extracted by solving Eq.~\eqref{Eq:SM: Approximation of edge transition equation} and is given by:
\begin{equation}
    \Tilde{\Delta}_{c} = \sqrt{-\frac{0.177715 W(-\alpha \log (| \delta \theta | ))}{\log (| \delta \theta | )}},
    \label{Eq:SM: Analytic expression for semiclassical edge transition}
\end{equation}
where $W$ denotes Lambert's W function, $\alpha = 0.000034$ and $\Tilde{\Delta}_{c}$ corresponds to the value of $\Tilde{\Delta}$ at which the edge transition occurs. We have used Eq.~\eqref{Eq:SM: Analytic expression for semiclassical edge transition} to plot the edge transition curve in Fig.~\ref{Fig:SM:Semiclassical edge transition} (inset).

For comparatively large values of $|t-t^{\prime}|$ we obtain an edge transition at $\Delta = E_{\rm{C}}$, since the second factor in Eq.~\eqref{SM:Eq: ratio of energy for charge and spin edge modes} equals one, which resembles the value of edge transition obtained using the perturbative calculation in Sec:\ref{SM:Sec:Perturbative calculation}.

\section{Field theory perturbing about the superconducting phase} 
\label{SM:Sec:Integrating out fermions}
The effect of the unitary transformation on the various operators in our model is given by:
\begin{gather}
\begin{aligned}
    U^{\dagger}\hat{N}_{X}U &= \hat{N}_{X} - \frac{\hat{n}_{X}}{2}, \\
    U^{\dagger}d_{j,\sigma,X}U &= d_{j,\sigma,X}e^{-i\frac{\hat{\phi}_{X}}{2}}.
\end{aligned}
\label{SM:Eq:The effect of unitary transformation}
\end{gather}
The transformed Hamiltonian ($\hat{H}^{\prime}$) is given by:
\begin{gather}
 \begin{aligned}
   \hat{H}^{\prime} &= E_{\rm{C}}\sum_{X} (2\hat{N}_{X} - N_{g})^{2} \\
   &-t\sum_{X,\sigma} (d^{\dagger}_{X,\sigma,A}d_{X,\sigma,B} + \mathrm{H.c.})\\
   &-t^{\prime}\sum_{X,\sigma} (d^{\dagger}_{X+1,\sigma,A}e^{\frac{i\hat{\phi}_{X+1}}{2}}d_{X,\sigma,B}e^{\frac{-i\hat{\phi}_{X}}{2}} + \mathrm{H.c.}) \\
   &-\frac{\Delta}{2}\sum_{X,j,\sigma,\sigma^{\prime}} (d^{\dagger}_{X,j,\sigma}[\sigma_{y}]_{\sigma,\sigma^{\prime}}d^{\dagger}_{X,j,\sigma^{\prime}} + \mathrm{H.c.}).
 \end{aligned}
\label{Eq: Hamiltonian after unitary transformation.}
\end{gather}
The transformed Hamiltonian also possesses the symmetries of the original Hamiltonian although the symmetry operations are modified. The sublattice/particle hole  transformation in the  new basis is given by:
\begin{equation}
    \mathit{C}^{\prime} = \mathit{C} e^{i\pi\sum_{X} \frac{\hat{n}_{X}}{2}}.
\end{equation}
Similarly, it can be seen that the transformed Hamiltonian also possesses the total charge symmetry of the original Hamiltonian. \\
The action corresponding to the transformed Hamiltonian (as highlighted in the methods section of the main text) is given by:
\begin{gather}
    \begin{aligned}
        S &= S_{0}(\phi) + S_{0}(d,\Bar{d}) + \Delta S_{1}(\phi,d,\Bar{d}) \\
        &+ \Delta S_{2} (\phi,d,\Bar{d}) ,
    \end{aligned}
\end{gather}
where,
\begin{small}
\begin{gather}
    \begin{aligned}
         S_{0}(\phi) &= \int_{x,\tau}\frac{\Dot{\phi}^{2}}{16E_{\rm{C}}a} ,\\
        S_{0}(d,\Bar{d}) &=\int_{x,\tau} \Bar{d}_{j}(x)\partial_{\tau}d_{j}(x)
        - t  (\Bar{d}_{A,\sigma}(x)d_{B,\sigma}(x) + \mathrm{H.c.})
\\ &-t^{\prime}(\Bar{d}_{A,\sigma}(x+a)d_{B,\sigma}(x) + \mathrm{H.c.})\\
&- \frac{\Delta}{2} (\Bar{d}_{j,\sigma}(x)[\sigma_{y}]_{\sigma,\sigma^{\prime}}\Bar{d}_{j,\sigma}(x) + \mathrm{H.c.}), \\
\Delta S_{1}(\phi,d,\Bar{d}) &= -\frac{t^{\prime}}{2} \int_{x,\tau}( i(\phi(x+a) - \phi(x))\Bar{d}_{A,\sigma}(x+a)d_{B,\sigma}(x) \\&+ \mathrm{H.c.}), \\
\Delta S_2(\phi,d,\Bar{d}) &= \frac{t^{\prime}}{8}\int_{x,\tau} ((\phi(x+a) - \phi(x))^{2}\Bar{d}_{A,\sigma}(x+a)d_{B,\sigma}(x) \\&+ \mathrm{H.c.}),\\
\label{Eq: Gradient expansion in phi}
    \end{aligned}
\end{gather}
\end{small}
where the fermionic variables $d/\Bar{d}$ represent Grassmann variables and $j \in \{A,B\}$. \\

\section{DMRG: Additional Figures and details on data extraction}
\label{Sec:DMRG and Data extraction}
\begin{figure}
    \includegraphics[scale = 0.07]{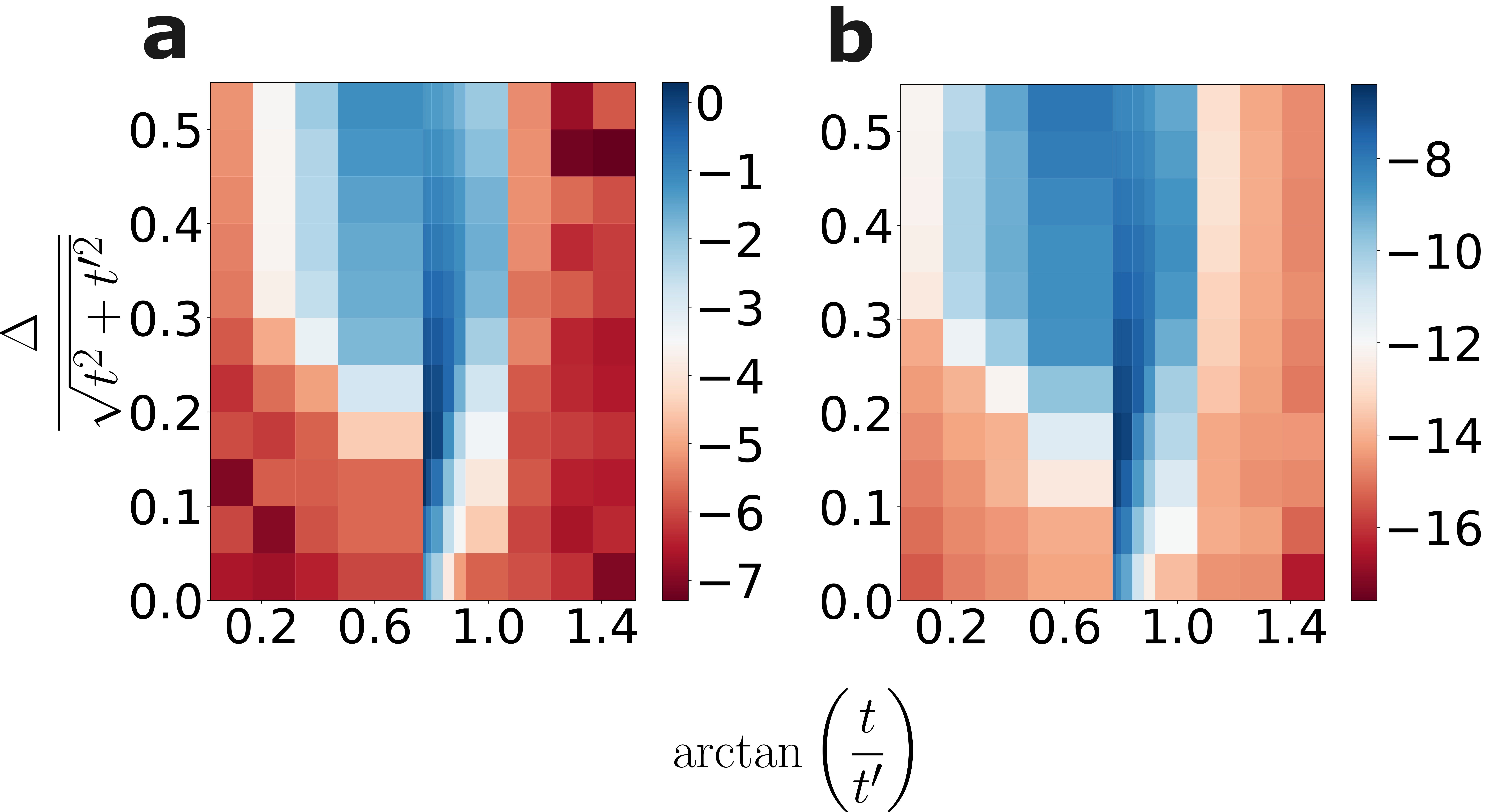}
    \caption{(a) Color plot of $\log_{10}$(Energy Variance) of the obtained ground state as function of model parameters. (b) Color plot of $\log_{10}$ (Maximum truncation error) of the obtained groundstate as a function of model parameters.}
    \label{SM:Fig:Variance and Maximum truncation error}
\end{figure}

\begin{figure}
    \vspace{2.5 pt}
    \centering
    \includegraphics[scale = 0.08]{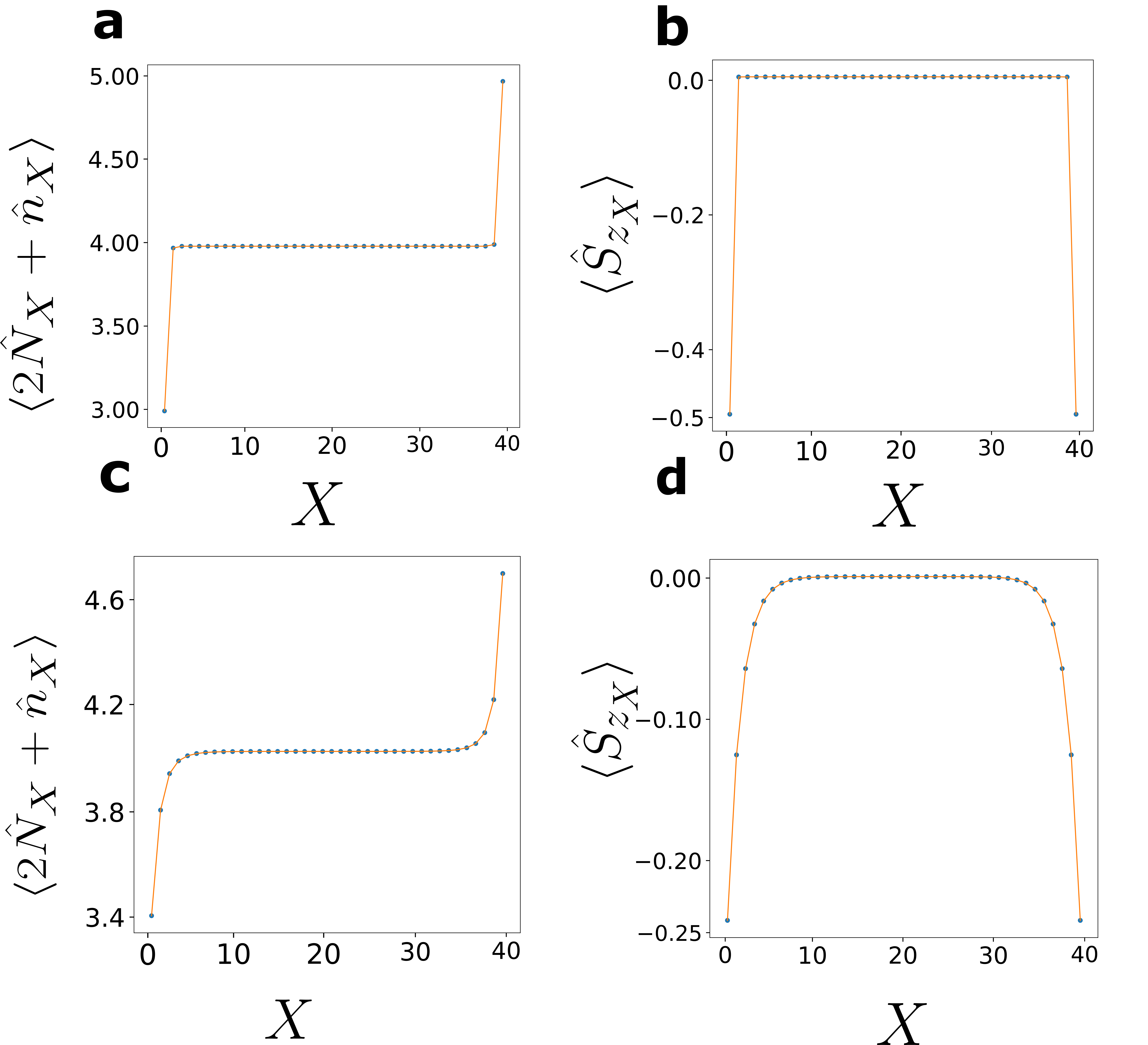}
    \caption{a) $\&$ b) Charge and spin edge states obtained for $\arctan (\frac{t}{t^{\prime}}) = 0.02$ for $\frac{\Delta}{E_{\rm{C}}} = 1.2$ and $0.2$ respectively.  c) $\&$ d)   Charge and spin edge states obtained for $\arctan (\frac{t}{t^{\prime}}) = 0.62$ for $\frac{\Delta}{E_{\rm{C}}} =1.6$ and  $0.4$ respectively. All plots are obtained for a system size of 40 and $N_{g} = 4$.}
    \label{SM:Fig:Edge states obtained using DMRG.}
\end{figure}

We discuss in this section the variance and truncation error corresponding to the numerically obtained phase diagram in Fig~\ref{Fig:DMRG}(a), numerically obtained edge state and their transition, arguments for the relevant error in the entanglement spectrum values and how we used it to obtain Fig~\ref{Fig:DMRG}(c) and the entanglement spectrum in the presence of symmetry breaking perturbations.\\
The logarithmic plot of the variance and maximum truncation error of the groundstate obtained using DMRG is shown in Fig.~\ref{SM:Fig:Variance and Maximum truncation error}. 
We suspect that the algorithm got stuck in a local minima in the regime corresponding to $t \approx t^{\prime}$ and small values of $\Delta$ since these state have relatively high variance (compared to states obtained deep in the insulating phases), of the order 1, while still exhibiting a truncation error of the order of $10^{-7}-10^{-8}$. The results remained unchanged for different initial states and relatively higher bond dimensions (all the results in this paper were obtained for a maximum bond dimension of 1000).\\
\begin{figure}[!htp]
    \centering
    \includegraphics[scale = 0.08]{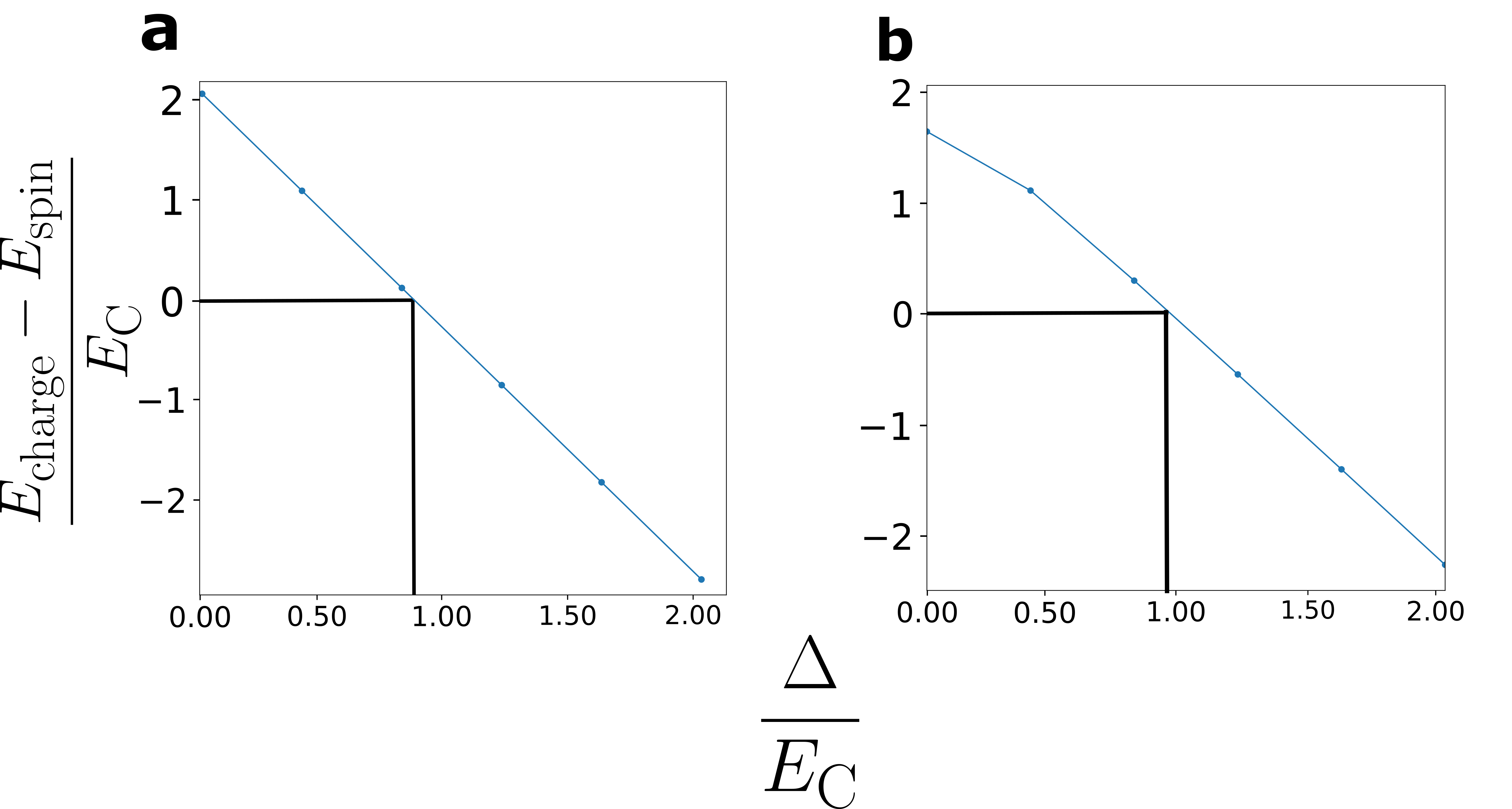}
    \caption{The difference in energy of charge  and spin edge modes as a function of $\frac{\Delta}{E_{\rm{C}}}$ for $\arctan (\frac{t}{t^{\prime}}) = 0.02$ (a) and $\arctan (\frac{t}{t^{\prime}}) = 0.62$ (b). All the plots are obtained for a system size of 40 and $N_{g} = 4$. 
    }
    \label{SM:Fig:Edge transition obtained using DMRG.}
\end{figure}
 The numerically obtained edge transition is shown in Fig.~\ref{SM:Fig:Edge transition obtained using DMRG.} for $\arctan(\frac{t}{t^{\prime}}) =  0.02 \hspace{2 pt} \& \hspace{2 pt} 0.62$. The results agree with the analytical results for a large regime of the phase diagram. It was not possible to study the behavior of edge transition near the phase transition due to numerical inaccuracy of DMRG results as explained above. The plot of edge states as a function of position is shown in Fig.~\ref{SM:Fig:Edge states obtained using DMRG.}. The integral of the corresponding curves is quantized upto numerical precision as expected based on our analytical results.

To obtain the entanglement spectrum plot in Fig.~\ref{Fig:DMRG}, we used the inbuilt function in the TENPY library. The error bars of the obtained spectrum values was calculated using the norm error obtained from the iDMRG run and is given by:
\begin{equation}
    \delta{\epsilon}_{i} = \frac{2\sqrt{\text{norm error}}}{\sqrt{e^{-\epsilon_{i}}}}.
    \label{SM:Eq: Entanglement_spectrum_error}
\end{equation}
We have not explicitly shown these error bars in Fig.~\ref{Fig:DMRG}(c) since we assigned the same value to all the values that were within the error bar of a given spectrum value $\epsilon_{i}$.

\begin{figure*}
\vspace{2.5 pt}
    \centering
    \includegraphics[scale = 0.3]{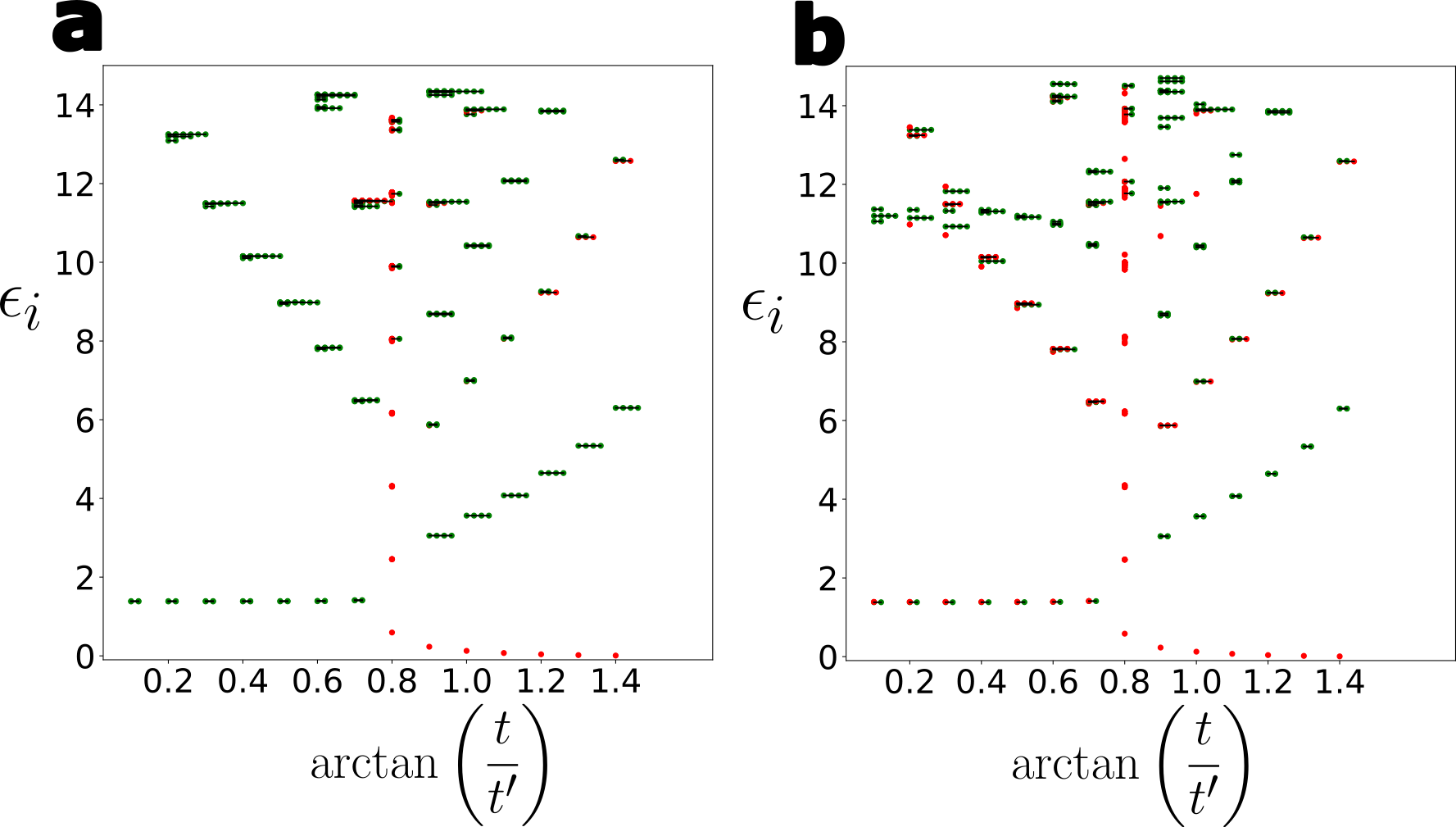}
    \caption{a) $\&$ b)  Entanglement spectrum when $\mathit{C}$ and $U(1)$ breaking terms respectively are added to the Hamiltonian of our model, Eq.~\eqref{Eq: Hamiltonian for the total model.}. Red dots indicate oddfold degenerate values while green dots represent evenfold degenerate values. As can be seen in a), breaking  $U(1)$ symmetry leads to degeneracy lifting in the SPT phase. Note that lowest even degenerate values (in the presence of $U(1)$ symmetry, see Fig.~\ref{Fig:DMRG} c)) break into even + odd + odd values in presence of $U(1)$ symmetry breaking term in b). The two odd values (shown in red) appear on top of each other in b) but are numerically different and can be distinguished using the error bound defined in Eq.~\eqref{SM:Eq: Entanglement_spectrum_error}. Note that we only show the entanglement spectrum values upto $\epsilon_{i} = 15$.}
    \label{SM:Fig:Entanglement for symmetry breaking perturbations}
    
\end{figure*}

To study the effect of symmetry breaking perturbations on the entaglement spectrum, we used the following terms $H_{\mathit{C}}$ and $H_{U(1)}$ as perturbations in our iDMRG simulations:
\begin{gather}
\begin{aligned}
     H_{\mathit{C}} &= \delta \sum_{X} (e^{-i\hat{\phi}_{X}}(d^{\dagger}_{A,X,\uparrow}d^{\dagger}_{B,X,\downarrow} + d^{\dagger}_{A,X,\downarrow}d^{\dagger}_{B,X,\uparrow})+ \mathrm{H.c.}),\\
     H_{U(1)} &=\delta \sum_{X} (d^{\dagger}_{A,X,\uparrow}d^{\dagger}_{B,X,\downarrow} + d^{\dagger}_{A,X,\downarrow}d^{\dagger}_{B,X,\uparrow}+ \mathrm{H.c.}),
\end{aligned}
\end{gather}
where $\delta = 0.1 E_{\rm{C}}$ for the iDMRG simulations. Note that since $\mathit{C}$ is an exact symmetry only for $N_{g} = 2$, we studied the effect of symmetry breaking perturbations for $N_{g} = 2$. The correspondong entanglement spectrum is shown in Fig.~\ref{SM:Fig:Entanglement for symmetry breaking perturbations}.
\newpage

\end{document}